\documentclass[aps, prd, twocolumn, showpacs, floatfix, letterpaper,
 nofootinbib, superscriptaddress,longbibliography]{revtex4-2}
\usepackage[version=4]{mhchem}
\usepackage{blindtext}
\usepackage[newitem]{paralist}
\usepackage{graphicx}
\usepackage{epsfig}
\usepackage{bm}
\usepackage{amsfonts}
\usepackage{pgf}
\usepackage{color}
\usepackage[T1]{fontenc}
\usepackage[latin1]{inputenc}
\usepackage{graphicx}
\usepackage[english]{babel}
\usepackage{amsmath}
\usepackage{amssymb}
\usepackage{amsfonts}
\usepackage{makecell}
\usepackage{hyperref}
\usepackage[draft,deletedmarkup=xout]{changes}
\usepackage{mathtools}
\usepackage{xcolor}
\usepackage{multirow}
\usepackage{mhchem}
\usepackage{wrapfig}
\usepackage{lipsum}
\usepackage{xcolor}
\usepackage{multirow}
\usepackage{mhchem}
\definecolor{green}{rgb}{0.19,0.64,0.54}
\definecolor{blue}{rgb}{0,0,1}
\definecolor{reddish}{rgb}{0.65, 0.2, 0.2}
\definecolor{darkgreen}{rgb}{0.2,0.7,0.3}
\definecolor{darkblue}{rgb}{0.3,0.40,0.48}
\definecolor{gray}{rgb}{.8,.8,.8}

\newcommand{\gem}[0]{\Gamma^{\text{em}}}
\newcommand{\yl}[0]{\mathcal{Y}}
\newcommand{\ogw}[0]{\Omega_{\rm gw}}
\newcommand{\rd}[0]{{\rm rd}}
\newcommand{\te}[0]{{\tilde e}}
\newcommand{\ggr}[0]{\Gamma^{\text{gr}}}

\hypersetup{,
	pdftitle={SGWB for chiral superconducting cosmic strings},
  	pdfauthor={I.Yu. Rybak and L. Sousa},
    colorlinks=true,       
    linkcolor=darkblue,          
    citecolor=darkgreen,        
    filecolor=reddish,      
    urlcolor=blue,          
    linktocpage=true
}

\begin{document}

\title{Stochastic Gravitational Wave Background from \\ Chiral Superconducting Cosmic Strings}

\author{I. Yu. Rybak}
\email[]{irybak@unizar.es}
\affiliation{CAPA \& Departamento de F\'{\i}sica Te\'{o}rica, Universidad de Zaragoza, Pedro Cerbuna, 12, 50009 Zaragoza, Spain}
\affiliation{Centro de Astrof\'{\i}sica da Universidade do Porto, Rua  das Estrelas, 4150-762 Porto, Portugal}
\affiliation{Instituto de Astrof\'{\i}sica e Ci\^encias do Espa\c co,  CAUP, Rua das Estrelas, 4150-762 Porto, Portugal}

\author{L. Sousa}
\email{Lara.Sousa@astro.up.pt}
\affiliation{Centro de Astrof\'{\i}sica da Universidade do Porto, Rua das Estrelas, 4150-762 Porto, Portugal}
\affiliation{Instituto de Astrof\'{\i}sica e Ci\^encias do Espa\c co, CAUP, Rua das Estrelas, 4150-762 Porto, Portugal}
\affiliation{Departamento de F\'{\i}sica e Astronomia, Faculdade de Ci\^encias, Universidade do Porto, Rua do Campo Alegre 687, PT4169-007 Porto, Portugal}

\begin{abstract}
We investigate the emission of vector radiation by superconducting cosmic string loops, deriving general relations to characterize the vector radiation emission efficiency, and study its impact on the evolution of loops. Building on these results, we compute the stochastic gravitational wave background generated by a chiral superconducting cosmic string network, including the impact of vector radiation for the very first time. Our analysis reveals that strong coupling between superconducting cosmic strings and the vector field may lead to a substantial suppression of the gravitational wave signal, while moderate coupling may still produce a detectable signal. We demonstrate that, in this intermediate limit, the presence of superconductivity in cosmic strings may help reconcile their gravitational wave spectrum with pulsar timing array data for large enough values of current.
\end{abstract}

\date{\today}
\maketitle

\section{Introduction}

Many high-energy physics scenarios predict the formation of one-dimensional topological defects known as cosmic strings in phase transitions in the early universe \cite{Kibble, HindmarshKibble,Vilenkin:2000jqa, JeannerotRocherSakellariadou,BhattacharjeeSahuYajnik,Allys,SarangiTye, Dror:2019syi}. Once cosmic strings are produced, they form a network that spans the entire universe. When cosmic strings collide, they intercommute and may form closed strings with lengths smaller than the horizon. These string configurations are called loops, while those that cross the horizon are known as long strings. Because of the highly nonlinear nature of cosmic strings, understanding their evolution typically requires numerical simulations. However, to date, only the simplest realizations of cosmic strings have been studied numerically in detail: local Abelian-Higgs U(1) models with critical coupling \cite{Hindmarsh:2008dw,HindmarshLizarragaUrrestillaDaverioKunz, CorreiaMartins,Correia:2021tok} and their infinitely thin approximation, known as Nambu-Goto simulations \cite{Martins:2005es,Ringeval:2005kr,Blanco-PilladoOlumShlaer}.

When cosmic strings are coupled to other fundamental fields, they can acquire superconductivity \cite{Davis:1988ij, Peter:1992ta, Davis:1995kk, Bin_truy_2004, Allys2, AbeHamadaYoshioka, FukudaManoharMurayamaTelem}. Although some progress has recently been made in simulating superconducting cosmic strings \cite{Battye:2021kbd, Hiramatsu:2023epr, Fujikura:2023lil, Correia:2024wsq, Battye:2024dvw}, the additional degrees of freedom associated with the current propagating on cosmic strings make numerical simulations particularly challenging. To address superconducting string networks, given the lack of full-scale network simulations, one can resort to a semi-analytical approach to describe their large scale evolution known as the Charge-Velocity-dependent One-Scale (CVOS) model, which has been developed in a series of recent studies \cite{MPRS,MPRS2,MPRS3}. While the CVOS model contains some model-dependent parameters that should be calibrated with future simulations, it also predicts generic features that should hold for all superconducting string networks.

The existence of cosmic strings potentially leads to many observational signatures that may allow us to detect these one-dimensional topological defects for the very first time. By investigating the existence of superconductivity on cosmic strings, we can probe potential interactions between fundamental fields and, even if cosmic strings are not directly detected, their absence should enable us to constrain different string-forming models for the early universe. Cosmic strings, therefore, serve as a valuable source of information, aiding in uncovering high-energy physics. Particularly sensitive probes of cosmic string scenarios may be conducted by studying the Cosmic Microwave Background (CMB) \cite{LizarragaUrrestillaDaverioHindmarshKunz, LazanauShellard, CharnockAvgoustidisCopelandMoss}, Gravitational Wave (GW) emissions \cite{LISA,NANOGrav:2023hvm}, lensing effects \cite{Sazhin1, Sazhin2}, and structure formation \cite{Jiao:2023wcn, Jiao:2024rcr}. In the case of superconducting cosmic strings coupled with the electromagnetic field, one can anticipate additional constraints from radiation channels \cite{Imtiaz:2020igv, Brandenberger:2019lfm, Cyr:2023yvj}. However, if the current on cosmic strings is related to the dark sector \cite{Hyde:2013fia}, electromagnetic constraints cannot be relied upon, leading to modifications in certain observational outcomes
 \cite{Long:2014mxa, Long:2014lxa}.

 The Stochastic Gravitational Wave Background (SGWB) produced by current-carrying cosmic strings was initially studied in Refs.~\cite{Auclair:2022ylu,Paper1}, but these studies did not account for the emission of vector radiation by cosmic string loops. In this study, we build upon on our previous work, as reported in Ref.~\cite{Paper1}, and perform more precise predictions for this background, by considering the impact of this vector radiation for the first time. In particular, we focus on studying the impact of the potential emission of vector radiation by current-carrying loops and on establishing phenomenological relations to describe the vector radiation emission efficiency. We present refined equations for the evolution of cosmic string loops, taking into account the radiation associated with current and possible current leakage effects. We study how these additional physical mechanisms influence the SGWB generated by a network of current-carrying strings. Furthermore, we conclude by discussing possible future research directions aimed at enhancing our understanding of current-carrying strings.

Throughout this paper, Latin indices ``$a$-$d$'' run over worldsheet coordinates (from $0$ to $1$), while Greek indices span over spacetime coordinates $X^{\mu}$ (from $0$ to $3$).

\section{Dynamics of infinitely thin current-carrying cosmic strings}
\label{Infinitely thin current-carrying cosmic strings}

We begin our analysis by setting up an effective action to describe superconducting cosmic strings. Here, we use the generalized Nambu-Goto action, assuming that the cosmic strings are infinitely thin and are coupled to a gauge vector field, as proposed in Refs.~\cite{Witten84, Peter:1992dw}. In this case, cosmic strings trace a 1+1 dimensional worldsheet in spacetime $X^\mu=X^\mu(\sigma^a)$, where $\sigma^a = \left\{ \tau, \sigma \right\}$ are worldsheet coordinates. The action is written as follows
\begin{equation}
\label{EffectiveAction}
S_{\text{w}} = S_{\text{eff}} + S_{\text{int}} + S_{\text{em}},
\end{equation}
where 
\begin{equation}
\label{Eff}
S_{\text{eff}} = \mu_0 \int \sqrt{-\gamma} \mathcal{L}(\kappa) d^2 \sigma,
\end{equation}
describes the effective action of an elastic string, $\mu_0$ is a constant defined by the symmetry breaking scale, $\gamma=\det (\gamma_{ab}) $, $\gamma_{ab} \equiv X^{\mu}_{,a} X^{\nu}_{,b} \eta_{\nu \mu}$ is the induced metric, where $\eta_{\mu \nu}$ is the Minkowski metric, and $X^{\mu}_{,a} \equiv {\partial X^{\mu}}/{\partial \sigma^{a}}$; 
\begin{equation}
\label{EM}
S_{\text{em}} = \frac{1}{16 \pi} \int \sqrt{-g} F_{\mu \nu} F^{\mu \nu} d^4 x,
\end{equation}
with $F_{\mu \nu} \equiv A_{\nu, \mu} - A_{\mu,\nu} $, is the action of the vector fields; and
\begin{equation}
\label{In}
S_{\text{int}} = e \int A_{a} \varepsilon^{ab} \phi_{,b} \, d^2 \sigma,
\end{equation}
describes the coupling between the current carriers and the vector fields, $e=\mu_0^{1/2} \tilde{e}$, $\tilde{e}$ is the charge of the current carriers~\footnote{Note that, for convenience and in contrast to other literature on cosmic strings \cite{Witten84, VilenkinVachaspati1987,MartinsShellard1998}, we use rescaled definitions of the field and coupling: specifically, $\phi = \tilde{\phi}/ \mu_0^{1/2} $, and $ e =  \tilde{e} \mu_0^{1/2}$, where tildes indicate the notation used in those references.}, $A_{a} \equiv A_{\mu} X^{\mu}_{,a}$ and $\varepsilon^{ab}$ is the Levi-Civita symbol. One may notice that the gauge invariance $A_{\mu} \rightarrow A_{\mu} + \partial_{\mu} \phi $ of the action (\ref{EffectiveAction})  is guaranteed by $\varepsilon^{ab}$ since  $\phi_{,a} = \phi_{,\mu} X^{\mu}_{,a} $. 

 The master function $\mathcal{L}(\kappa)$ has an internal degree of freedom, the 4-current, given by
\begin{equation}
    \label{kappa}    \kappa \equiv \phi_{,a} \phi_{,b} \gamma^{ab}, 
\end{equation}
where $\phi$ is a scalar field describing the charge carriers that are confined to the worldsheet. This function defines the properties of the current-carrying strings (see Ref.~\cite{Carter2} for a detailed review). Particularly, by defining the form of the function $\mathcal{L}(\kappa)$, one determines how the mass per unit length and tension (or equivalently, how the speed of the longitudinal and of the transverse perturbations) depend on the value of the current $\kappa$.  There are two conditions that we may impose on $\mathcal{L}(\kappa)$ to obtain wave-like equations of motion for the cosmic string, that admits exact solutions in terms of left- and right-moving modes (labelled respectively by the subscripts `$+$' and `$-$') for the string position
 \begin{equation}
 \label{StringSol}
      X^{\mu}(\tau, \sigma) = \frac{1}{2} \left( X^{\mu}_+(\sigma_+) + X^{\mu}_-(\sigma_-) \right) 
 \end{equation}
and for the scalar field 
\begin{equation}
\label{PhiSol}
    \phi(\tau, \sigma) = \frac{1}{2} \left( F_+(\sigma_+) + F_-(\sigma_-)  \right),
\end{equation}
where $\sigma_{\pm} = \tau \pm \sigma$. In this case, $F'_\pm$ may be regarded, in a sense, as left- and right-moving currents (while the current on the string is still given by Eq.~\eqref{kappa}). The first condition is the so called chiral limit, reached when $\kappa \rightarrow 0$ \cite{Carter:1999hx, Blanco-Pillado:2000ssd, Davis:2000cx}. In this case, the current $\kappa$ is a light-like field on the string world-sheet, which implies that either the right- or left-mover of solution \eqref{PhiSol} vanishes: $F_+=0$ or $F_- = 0$. Thus, for any $\mathcal{L}(\kappa)$, the solution of the equations of motion for the cosmic string in the chiral limit ($\kappa \rightarrow 0$) is given by \eqref{StringSol} and \eqref{PhiSol}, while the norm of $X^{\prime \, \mu}_{\pm}$ is related to the scalar field through \cite{Rybak:2018oks}
\begin{equation}
    \label{ChiralSol}
    X^{\prime \, \mu}_{\pm} X^{\prime}_{\pm \, \mu} = - 2  F^{\prime \, 2}_{\pm}  \frac{d\mathcal{L}(\kappa)}{d \kappa} \Bigg|_{\kappa = 0}.
\end{equation}
Another possibility is to demand that the string is transonic --- i.e. that the  longitudinal and transverse perturbations move at the same speed. This condition leads to $\mathcal{L}(\kappa) = \sqrt{1-\kappa}$ and also implies that the general solution is given by \eqref{StringSol} and \eqref{PhiSol} \cite{Carter:1990nb} and that the relation between $X^{\prime \, \mu}_{\pm}$ and the scalar field is given by Eq.~\eqref{ChiralSol} as well \cite{Rybak:2020pma}.

Here, by taking advantage of the integrability of their equations of motion, we will use chiral and transonic strings as a proxy to study the vector field radiation emitted by current-carrying cosmic strings. Although superconducting strings are not generally expected to always be chiral or transonic (see e.g.~\cite{Carter:1997pb}), as we shall see, there are reasons to believe that our findings should apply to any type of current-carrying cosmic string.

\section{Emission of vector radiation by current-carrying loops}\label{sec:vectorrad}

Here, we investigate the vector radiation emitted by superconducting string loops. These loops oscillate with a period of $T=L/2$, where $L$ is a constant that coincides with the loop length $\ell$ when the current is absent (further details about this relation are provided in Sec.~\ref{osc}), and are expected to emit vector radiation in a discrete set of frequencies determined by harmonics of $L$: $\omega_j=4\pi j/L$, where $j$ represents the harmonic number, and $\omega_j$ denotes the corresponding frequency. The power radiated in the direction of the unit vector $\textbf{n}$ per unit solid angle $\Omega(\theta, \varphi)$, averaged over a loop period $T$, may be expressed as follows~\cite{VilenkinVachaspati1987}:
\begin{equation}
\begin{gathered}
    \label{PowerRad}
  P = \sum_{j} P_j = e^2 \gem, \\ \frac{d P_j}{d \Omega} = - \frac{\omega_j^2}{2 \pi} j^{\mu} j^{*}_{\mu} = e^2 \frac{d\gem }{d\Omega}\,,
\end{gathered}
\end{equation}
where the subscript `$j$' indicates that we are considering the contribution of the $j$-th harmonic mode of the corresponding variable,
\begin{equation}
\begin{gathered}
\label{n}
\textbf{n} = \left( \cos \theta, \; \sin \theta \cos \varphi, \; \sin \theta \sin \varphi \right) \,, \\
\end{gathered}
\end{equation}
and we have defined
\begin{equation}
\begin{gathered}
    \label{IJpm}
    j^{\mu} = e \frac{L}{8 \pi^2} \left( I_- J_+^{\mu} - I_+ J_-^{\mu} \right), \\
    I_{\pm} = \int_0^{2 \pi}  F^{\prime}_{\pm} \text{e}^{ i j \left( \sigma_{\pm} - \frac{2 \pi}{L} \textbf{n} \cdot \textbf{X}_{\pm} \right) } d \sigma_{\pm}, \\
    J_{\pm}^{\mu} = \int_0^{2 \pi} X^{\prime \, \mu}_{\pm} \text{e}^{ i j \left( \sigma_{\pm} - \frac{2 \pi}{L} \textbf{n} \cdot \textbf{X}_{\pm} \right) } d \sigma_{\pm}\,.
\end{gathered}
\end{equation}

Here, $\gem$ is the vector radiation emission efficiency --- or, in other words, the power emitted in units of $e^2$ --- which may be split into the contribution of each harmonic mode $\gem_j$. This is the crucial quantity one needs to compute to describe the emission of vector radiation by superconducting string loops. To do so we will assume $F^{\prime}_{\pm} = \text{constant}$ and consider two types of solutions:  Burden \cite{BURDEN1985} and Garfinkle-Vachaspati \cite{GarfinkleVachaspati2} loops. In this section, we only provide a brief outline of the results, but a detailed description of these loop solutions and the derivation of $\gem$ may be found in Appendixes~\ref{Burden's loop} and \ref{Cuspless loop}.

\begin{figure}[t]
\centering
		\includegraphics[width=\linewidth]{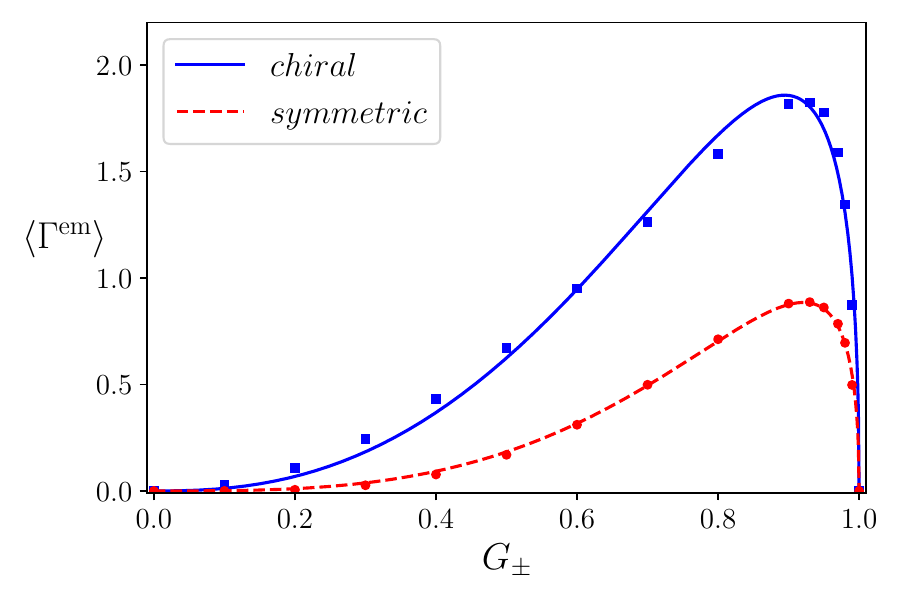}
\caption{Averaged vector radiation emission efficiency $\left\langle \gem \right\rangle$ of Burden loops for different values of $G_{\pm}$. Squares represent the values of $\left\langle \gem \right\rangle$ computed for loops with symmetrical currents ($G_{+}=G_-$), while circles correspond to those for loops with chiral currents ($G_{\pm}=1$, while $G_{\mp} \neq 1$). The solid (dashed) line represents the best fit for chiral (symmetric) currents, modeled by a function in the form of Eq.~(\ref{Gfun}). The best-fit parameter values are listed in Table \ref{TableAD}.}
\label{fig:BurdenAv}
\end{figure}

\begin{figure}[t]
\centering
		\includegraphics[width=\linewidth]{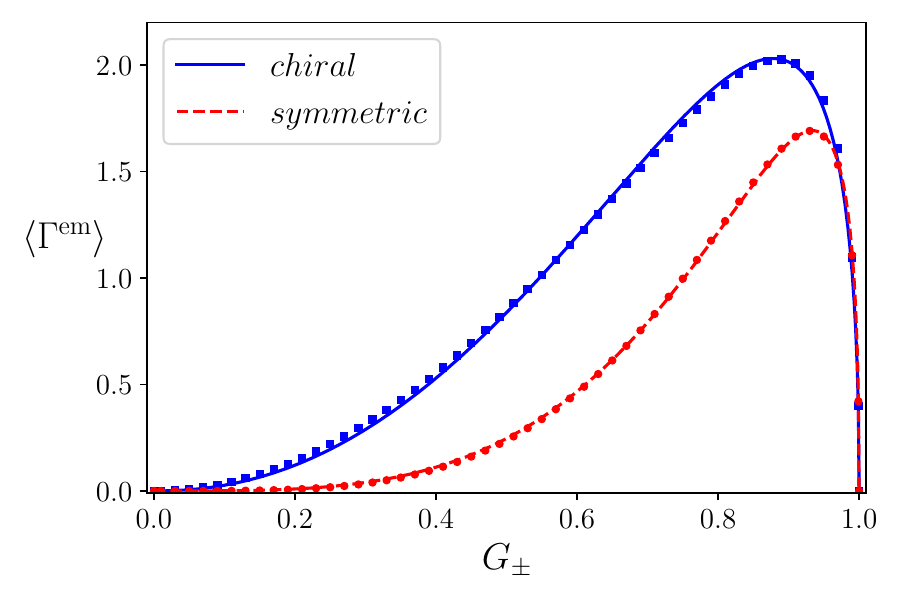}
\caption{Averaged vector radiation emission efficiency $\left\langle \gem \right\rangle$ of Garfinkle-Vachaspati loops for different values of $G_{\pm}$. Squares represent the values of $\left\langle \gem \right\rangle$ computed for loops with symmetrical currents ($G_{+}=G_-$), while circles correspond to those for loops with chiral currents ($G_{\pm}=1$, while $G_{\mp} \neq 1$). The solid (dashed) line represents the best fit for chiral (symmetric) currents, modeled by a function in the form of Eq.~(\ref{Gfun}). The best-fit parameter values are listed in Table \ref{TableAD}. }
\label{fig:CuspLessGeAv}
\end{figure}

In Appendix \ref{Burden's loop}, we derive the spectrum of emission $\gem_j$ for loops with quasi-cusp points using Eqs.~\eqref{PowerRad} and~\eqref{IJpm}, by resorting to the Burden loop solution~\cite{BURDEN1985}. Cusps are points on a string where the induced metric on the world-sheet becomes singular, causing these points to move at the speed of light. However, if the current-forming scalar field is non-trivial (i.e. $ F^{\prime}_{\pm}  \neq 0$) at these locations, the formation of cusps is prevented regardless of the choice of the master function $\mathcal{L}(\kappa)$ (a formal proof of this result is provided in Appendix \ref{NoCusps}). We define a quasi-cusp as a point that would form a cusp in the absence of current ($F^{\prime}_{\pm} = 0$). These points may exhibit shapes similar to conventional cusps but necessarily have subluminal velocities (see e.g.~\cite{Paper1}). Our analysis demonstrates that for loops with quasi-cusps, the efficiency of vector radiation emission decreases exponentially with increasing harmonic mode. This is a direct consequence of the absence of true cusps in superconducting strings. Consequently, when calculating $\gem$, the summation can be truncated at some maximum mode $j_{\rm max}$ without significant loss of precision. 

The resulting vector radiation emission efficiency, averaged over all possible Burden loop shapes (i.e. averaging over parameter $\beta$ in Appendix \ref{Burden's loop}), $\langle \gem \rangle$ as a function of current, expressed here, for simplicity, in terms of $G_{\pm} \equiv \sqrt{1- F^{\prime \, 2}_{\pm} }$, is displayed in Fig.~\ref{fig:BurdenAv}. The behavior of $\langle \gem \rangle$ is well described by a phenomenological
expression of the form
\begin{equation}
    \label{Gfun}
    \left\langle \gem \right\rangle = \gem_0 \left| F^{\prime}_{\pm} \right| \left( 1- \left| F^{\prime}_{\pm} \right| \right)^{D},
\end{equation}
where $\gem_0$ and $D$ are constant parameters. In Fig.~\ref{fig:BurdenAv}, one can see that this expression accurately describes the emission of vector radiation by Burden loops, using the fitting values listed in Table \ref{TableAD}.

This analysis shows that, although current is (obviously) essential for the emission of vector radiation, as current increases there is a suppression of the efficiency of emission. As a matter of fact, $\langle \gem \rangle$ peaks at moderate currents of roughly $F'_\pm \sim 0.4$. Moreover, while the results are qualitatively similar in both cases, the emission efficiency of vector radiation is higher for chiral currents than for symmetric currents.

For Garfinkle-Vachaspati loop solutions~\cite{GarfinkleVachaspati2}, which describe loops with kinks, we find a very similar scenario for the averaged vector radiation  emission efficiency $\langle \gem \rangle$ (we average over possible shapes controlled by parameter $\beta$ in Appendix \ref{Cuspless loop}) --- which we display in Fig.~\ref{fig:CuspLessGeAv}. Again $\langle \gem \rangle$ may be described by a phenomenological relation of the form in Eq.~\eqref{Gfun} albeit with different fitting constants (as shown in Table~\ref{TableAD}). The main difference between loops with quasi-cusps and loops with kinks may be seen in the spectrum of emission, whose computation is detailed in Appendix~\ref{Cuspless loop}. For loops with kinks, we found that $\gem_j$ follows a power-law, without the additional exponential suppression found for Burden loops. The dominant contribution to vector radiation in high frequencies for current-carrying strings is then generated by kinks, rather than quasi-cusps (similar to what was shown for GWs \cite{Paper1}).

Our results show that the vector radiation efficiency for both Burden loops with chiral and transonic currents and for Garfinkle-Vachaspati loops is well approximated by a relation of the form of \eqref{Gfun}. Note also that the current-carrying Garfinkle-Vachaspati loops we consider are a solution for the general superconducting string equations of motion --- i.e. for any master function $\mathcal{L}(\kappa)$ --- and so these results apply, in this case, to superconducting strings in general.  This suggests that this relation may, in fact, hold for any type of current-carrying loops (with different fitting constants).

\begin{table} [t]
\centering
\centering
\caption{Best fit values of the parameters of function~\eqref{Gfun}, for loops with quasi-cusp points and 
loops with kinks.}
\label{TableAD}
  \begin{tabular}{ | l | c | c |   }
    \hline
     & $\gem_0$  & $ D $  \\ \hline
    $\text{Symmetric loop with quasi-cusps}$ & $4.9$ & $1.6$   \\ \hline
     $\text{Symmetric loop with kinks}$ & $10.5$ & $1.8$   \\ \hline
    $\text{Chiral loop with quasi-cusps}$ & $8.6$ & $1.1$  \\ \hline
    $\text{Chiral loop with kinks}$ & $8.6$ & $1.2$   \\ \hline
  \end{tabular}
  \end{table}

\section{Cosmological evolution of networks of current-carrying strings}
\label{CVOS model}

Cosmic string loops are produced dynamically throughout the evolution of a cosmic string network as a result of (self-)intersections and subsequent intercommutation. Understanding the large-scale evolution of cosmic string networks is therefore essential to accurately compute the number density of loops (see e.g.~\cite{Sanidas:2012ee,Blanco-PilladoOlumShlaer,Kuroyanagi:2012wm,SousaAvelino}), which is an essential ingredient in the computation of their SGWB. Here, we will resort to the Charge-Velocity-dependent One-Scale (CVOS) model introduced in~\cite{MPRS} --- which we review in this section --- to describe the evolution of networks of cosmic strings with current.

\subsection{The CVOS model}\label{subsec:cvos}

The CVOS model  --- an extension of the VOS model~\cite{Martins:1996jp} to account for the dynamical impact of current ---  provides a thermodynamic description of the cosmological evolution of a current-carrying cosmic string network in a Friedman-Lema\^itre-Robertson-Walker (FLRW) universe, with line element
\begin{equation}
    \label{FLRW}
    ds^2 = a^2(\eta) (d \eta^2 - d \textbf{x}^2),
\end{equation}
where $a$ is the cosmological scale factor, $\eta$ is the conformal time and $\textbf{x}$ are comoving coordinates.

Let $\mathcal{F}(K)$ be an averaged equation of state for the network  --- obtained by averaging the worldsheet Lagrangian $\mathcal{L}(\kappa)$ over the whole network --- that depends on the macroscopic 4-current $K$, describing the averaged current of the strings in the network (see Refs.~\cite{Peter:1992dw,Carter:1997pb,MPRS} for a detailed description). The study in Ref.~\cite{MPRS3} demonstrated that a detailed modeling of the equation of state --- as carried out for the Witten $\rm U(1)\times U(1)$ model --- may be omitted in many situations and that the cosmological evolution of a network of current-carrying strings may be described approximately by considering a linear equation of state of the form $\mathcal{F}(K) = 1 - K/2$ studied in Ref.~\cite{MPRS2}. Here we will also follow this approach. 

In this model, the evolution of the cosmic string network is described by four macroscopic variables:

$\bullet$ $\xi$ - the characteristic length of the network, which is related to the bare string energy density (without current) through $\rho_0 = {\mu_0}/{ \xi^2}$;

$\bullet$ $v$ - the Root-Mean-Squared (RMS) velocity of the strings;

$\bullet$ $Y$ - the charge amplitude;

$\bullet$ $K$ - a macroscopic Lorentz-invariant current.

The linear CVOS model equations for the cosmological evolution of a network of current-carrying strings are given by~\cite{MPRS3}
\begin{subequations}
\label{EqOfMotMacroLin2} 
\begin{align}
    \label{EqOfMotMacroLinA2}    
    \frac{d \epsilon}{d x} =& \frac{\epsilon v^2}{1+Y} + \frac{c v}{2} + Y \frac{v k(v)}{1+Y} + \frac{\epsilon}{a H} \frac{d (a H)}{d x}, \\
    \label{EqOfMotMacroLinB2}
    \frac{d v}{d x} =& \frac{1-v^2}{1+Y} \left[ \frac{(1-Y) k(v)}{\epsilon} - 2 v \right], \\
    \label{EqOfMotMacroLinC2}
    \frac{d Y}{d x} =& 2 Y \left( \frac{v k(v)}{\epsilon} - 1 \right) - A(Y) \frac{Y}{\epsilon},
    \end{align}
\end{subequations}
 where, for simplicity, we have introduced new variables: $x= \log a$ and $\epsilon = \xi a H$. Moreover, $H=(da/dt)/a$ is the Hubble parameter, which evolves according to
 \begin{equation}
 H ^2 = H_0^2 \Omega_r \left( \frac{a_0}{a} \right)^4 + \Omega_m \left( \frac{a_0}{a} \right)^3 + \Omega_{\Lambda}  
 \end{equation}
where $t$ is the physical time (related to conformal time as $a d\eta = dt$), and the Hubble constant (the value of $H$ at the present time $t_0$) is given by $H_0 = 100 h \, \rm{km/s/Mpc}$, with $h = 0.678$. We take $a(t_0)\equiv a_0=1$ and the density parameters of radiation, matter, and dark energy at the present time are, respectively, given by $\Omega_r = 9.1476\times 10^{-5}$, $\Omega_m = 0.308$, and $\Omega_\Lambda = 1 - \Omega_r - \Omega_m $~\cite{Planck:2018vyg}. Note that the CVOS equations reduce to VOS equations if on sets $Y=0$.

\begin{widetext}
    
\begin{figure}[t]
\centering
        \includegraphics[scale=0.58]{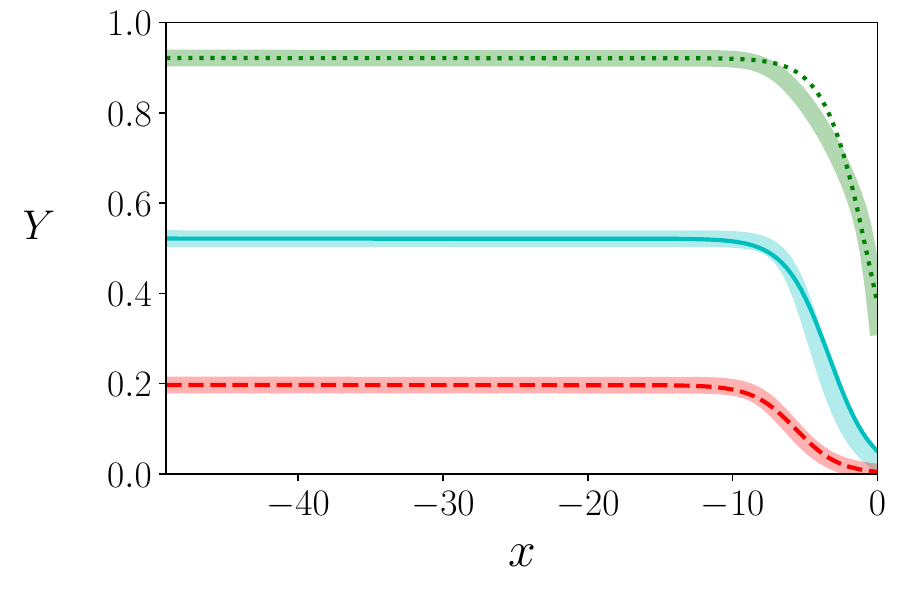} \includegraphics[scale=0.58]{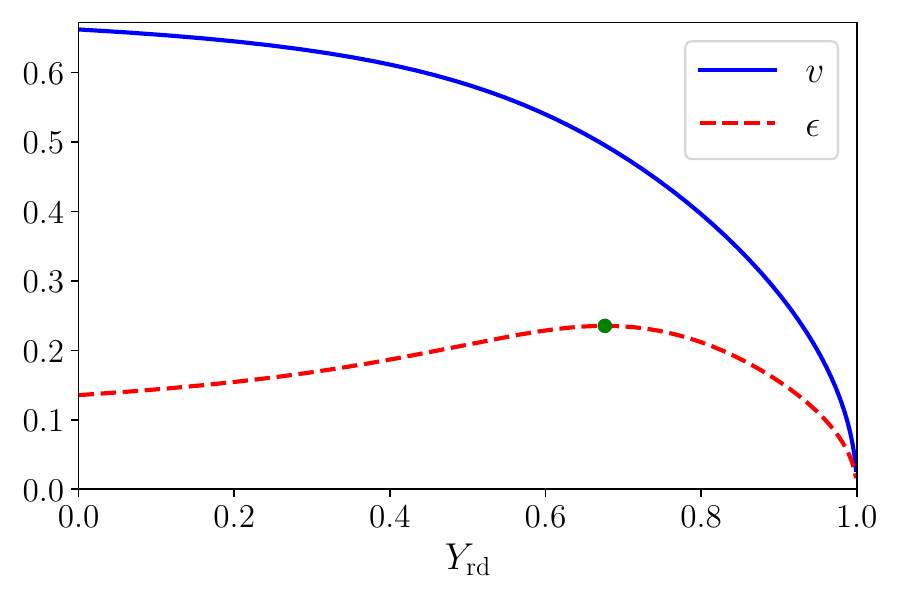}
\caption{ \label{fig:CVOS2} The left panel represents the evolution of the current amplitude $Y$ with ${Y}_{\rm cr}=1$, ${Y}_{\rm cr}=0.6$ and ${Y}_{\rm cr}=0.5$ when $A_{\rm const} = 10^{-3}$ (top to bottom). Thick transparent lines represent the evolution of $Y(x)$  in the model with leakage function \eqref{Af}, while thin dotted, solid and dashed lines demonstrate approximation \eqref{Yphen}.
The right panel represents the dependence of $v$ (solid line) and $\epsilon$ (dashed line) on the value of $Y_{\rm rd}$ during the radiation-dominated epoch, in which $d(aH)/dx=-aH$. The dot represents the maximum value of $\epsilon_{\rm max}=0.235$ achieved when $Y_{\rm rd}=Y_{\rm max} \approx 0.676$.}
\end{figure}
\end{widetext}

 
 The CVOS model \eqref{EqOfMotMacroLin2} contains two parameters , which we assume to take the same values as those in currentless Nambu-Goto cosmic string networks: $c=0.23$, representing the loop-chopping efficiency, and a momentum parameter given by \cite{Martins:2000cs}
\begin{equation}
    \label{Momentum}
    k(v)= \frac{2 \sqrt{2}}{\pi} \frac{1-8 v^6}{1+8 v^6} (1-v^2) (1+2 \sqrt{2} v^3).
\end{equation}
We did not include a decoupled equation for the macroscopic variable $K$, as it does not affect the evolution of the string network in the linear approximation, and we follow the model described in Ref.~\cite{MPRS3}.

Moreover, the linear CVOS model contains a charge leakage parameter $A(Y)$ that characterizes the efficiency of charge amplitude loss during the evolution of the network. A detailed description of the effect of $A(Y)$ on the evolution of the string network is provided in Ref.~\cite{MPRS2}. Here, we use the form proposed in \cite{MPRS3}:
\begin{equation}
\label{Af}
   A({Y}) = \frac{A_{\rm const}}{1-\text{e}^{-({Y}-{Y}_{\rm cr})^2}}, 
\end{equation} 
where $A_{\rm const}$ is a constant and ${Y}_{\rm cr}$ is the critical value of current above which charge leakage becomes highly efficient. This value is model-dependent and is determined by the mass of the string-forming Higgs field, $m_H$, and the mass of the current-generating condensate $m_{\sigma}$. For the Witten model, ${Y}_{\rm cr} = {2 m^2_{\sigma}}/{3 m^2_H}  $ \cite{Peter:1992dw}; however, we allow other values of the critical current to cover additional possible scenarios, imposing only the constraint that ${Y}_{\rm cr}  < 1$ . 

The evolution of a current-carrying string network with non-trivial charge leakage $A(Y)$ is characterized by a scaling solution, where $\epsilon$, $v$, and $Y$ asymptotically behave as constants for a fixed expansion rate (i.e for $a \propto t^{\lambda}$, with constant $\lambda$). Examples of the cosmological evolution of the charge amplitude $Y$ for such a network are displayed in Fig.~\ref{fig:CVOS2}. Therein, one may see that in the radiation-dominated epoch (which is well described by $\lambda = 1/2$), the network indeed experiences a linear scaling regime. However, as the universe transitions into the matter-dominated era, characterized by a faster expansion rate ($\lambda = 2/3$), the current amplitude on strings is dissipated.

\subsection{One parameter approximation\label{subsec:cvosapp}}

For simplicity, when studying the SGWB, it is convenient to reduce the number of parameters that control the value of the charge amplitude $Y$ within the CVOS model. As a non-trivial scaling of the charge occurs solely during the radiation-dominated epoch and it quickly declines for faster expansion rates \cite{MPRS2}, we may encapsulate the entire charge evolution by this value during the radiation-dominated era $Y_{\rm rd}$ and assume it decreases fast as expansion rate increases. Following a similar approach to that of Ref.~\cite{Rybak:2024djq}, we approximate the evolution of the charge amplitude as follows:
\begin{equation}
\label{Yphen}
    Y(x) = Y_{\rm rd} \left[\frac{1}{2} - \frac{1}{2}  \tanh\left( \frac{x+7(1-Y_{\rm rd})}{3} \right)   \right], 
\end{equation}
where the constants were chosen to reproduce the full model as closely as possible for a realistic cosmological background\footnote{Although the approximation in~\cite{Rybak:2024djq} may be used for different cosmological parameters, here we opted for developing a specific approximation for our fixed set of parameters (that come from Planck data) because this allows us to obtain a better fit for the realistic evolution of the network. Note that, as shown in~\cite{SousaAvelino}, having an accurate description of the large scale dynamics of the network is essential to perform accurate computations of the number of loops produced in their evolution and that any deviations may result in unphysical signatures on the SGWB.}. 

The evolution of the CVOS model is then reduced to the first two ordinary differential Eqs.~\eqref{EqOfMotMacroLinA2}-\eqref{EqOfMotMacroLinB2}, which then form a closed system with function \eqref{Yphen}. To verify the validity of this approximation, we evolve the full system of Eqs.~\eqref{EqOfMotMacroLin2} alongside the reduced one. The results are displayed on the left panel of Fig.~\ref{fig:CVOS2}, wherein one may see that this approximation provides a very good description for the evolution of $Y$, which necessarily translates into a good description of $\epsilon$ and $v$.
The phenomenological approximation \eqref{Yphen} is justified by the fact that the non-trivial charge amplitude $Y_{\rm rd}$ is present only in the radiation-dominated epoch, while it evolves toward zero for faster expansion rates. The only differences that appear between our approximation \eqref{Yphen} and the full model is in the radiation-matter transition period when the network is out of scaling, but, as this figure shows, these are minor. Since the charge amplitude $Y_{\rm rd}$ is solely controlled by the leakage function $A(Y)$, which depends on the underlying cosmic string model, and whose form we do not know even for the simplest realization $U(1)\times U(1)$ (due to a lack of numerical simulations), we assume Eq.~\eqref{Yphen}. By doing so, we transfer our lack of knowledge about $A(Y)$ to Eq.~\eqref{Yphen}, which should manifest itself only in the transition between the radiation and matter eras. Hence, we will study theoretically possible models rather than a particular realization of a superconducting cosmic string network.

It is useful to illustrate the dependency of the scaling values of $v$ and $\epsilon$ on the charge amplitude $Y_{\rm rd}$ during the radiation-dominated epoch, as these are essential ingredients in the computation of the number of loops produced. By identifying attractors for the reduced CVOS system and varying the values of $Y_{\rm rd}$, as depicted on the right panel of Fig.~\ref{fig:CVOS2}, we study how the evolution of the string network is influenced by the charge amplitude and, consequently, as the number of loops produced by the network varies. From Fig.~\ref{fig:CVOS2}, it becomes apparent that the RMS velocity monotonically decreases with an increase in the charge $Y_{\rm rd}$. Naively, this would mean that strings are less likely to collide to form loops as current increases. Meanwhile, however, variable $\epsilon$, which is inversely proportional to the density of the string network, exhibits a slight increase within the interval $0<Y_{\rm rd}<0.67$, reaching its maximum value  --- that we denote by $\epsilon_{\rm max}$ and is reached when $Y_\rd=Y_{\rm max}\approx 0.67$ ---  and subsequently decreases, approaching zero as $Y_{\rm rd} \rightarrow 1$. This behavior suggests that within the interval $0 < Y_{\rm rd} < 0.67$, changes in $\epsilon$ are primarily driven by the potential energy of strings, as determined by their curvature. However, for larger values of $Y_{\rm rd}$, the dominant mechanism behind changes in $\epsilon$ shifts to a reduction in energy loss caused by the decreasing RMS velocity $v$. As $Y_{\rm rd} \to 1$, the network becomes so dense that loop production is expected to become copious. These two phases were also identified in~\cite{Paper1} and have been examined in the context of predicting the CMB signatures arising from current-carrying cosmic strings~\cite{Rybak:2024djq}.

\section{Decay of loops due to gravitational and vector radiation}

The only thing missing to compute the SGWB generated by networks of current-carrying strings is to understand how loops with current decay by emitting gravitational and vector radiation and how much of their energy is emitted in the form of GWs. In this section, we will describe the decay of loops with current and establish a connection between the CVOS model and the properties of current-carrying loops.

\subsection{Relation between the network macroscopic variables and the microscopic variables for loops}

We will consider a simple model to describe the evolution of cosmic string loops in the presence of current, assuming that the loops are small when compared to the Hubble radius and thus neglecting the impact of cosmological expansion on their evolution.

Current on a cosmic string network is characterized by two main macroscopic parameters: a Lorentz scalar denoted by $K$ and the previously mentioned charge amplitude denoted by $Y$ (see Refs.~\cite{MPRS, MPRS2} for more details). Meanwhile, current on loops is characterized by the microscopic values $ F^{\prime}_{\pm} $. In this section, we will start by establishing a connection between the macroscopic variables describing the long string network and the microscopic values for oscillating loops.

Working within the assumptions of the CVOS model, our objective is to express $K$ and $Y$ as functions of $F^{\prime}_{\pm}$. These variables, however, appear in different worldsheet parametrizations. Specifically, in Sec.~\ref{Infinitely thin current-carrying cosmic strings} (and in Ref.~\cite{Rybak:2020pma}), a particular string world-sheet parametrization involving $(\tau, \sigma)$ is utilized, allowing the string solution to have a simple wavelike form. In contrast, the CVOS model employs a temporal-transverse gauge for the world-sheet parametrization, defined as $(t,s)$, which satisfies the following conditions:
\begin{equation}
    \label{TempGauge}
    \dot{ \textbf{X}} \cdot \textbf{X}^{\prime}_{s} = 0, \quad X^0 = t,
\end{equation}
where $\dot{\textbf{X}} \equiv {d \textbf{X}}/{d t}$ and $\textbf{X}^{\prime}_s \equiv {d \textbf{X}}/{d s}$.
In this gauge current can be expressed in the following form~\cite{MPRS, MPRS2}
\begin{equation}
 \kappa = \tilde{q}^2 - \tilde{j}^2, 
\end{equation}
where $\tilde{q}^2=\gamma^{00} \dot{\phi}^2$ and $\tilde{j}^2 = - \gamma^{11} \phi^{\prime \, 2} $. In the $(\tau,\sigma)$-parametrization, the 4-current can be expressed as follows \cite{Rybak:2020pma}:
\begin{equation}
    \label{CurrentPar}
    \kappa = \frac{4 F^{\prime}_{+} F^{\prime}_{-} }{1+ F^{\prime}_{+} F^{\prime}_{-}  - \textbf{X}^{\prime}_+ \cdot \textbf{X}^{\prime}_- } ,
\end{equation}
where we neglected the effect of the expansion of the universe and we have normalized linear solution as 
\begin{equation}
\phi^2 = \frac{1}{2} \left( F^{\prime}_{+} \sigma_+ + F^{\prime}_{-} \sigma_- \right)^2 .
\end{equation}
From Eq.~\eqref{CurrentPar}, one may deduce that 
\begin{equation}
\begin{gathered}
    \label{qj}
    \tilde{q}^2 = \frac{ \left( F^{\prime}_{+} + F^{\prime}_{-} \right)^2 }{1+ F^{\prime}_{+} F^{\prime}_{-} - \textbf{X}^{\prime}_+ \cdot \textbf{X}^{\prime}_- },\\
    \tilde{j}^2 = \frac{ \left( F^{\prime}_{+} - F^{\prime}_{-} \right)^2 }{1+ F^{\prime}_{+} F^{\prime}_{-} - \textbf{X}^{\prime}_+ \cdot \textbf{X}^{\prime}_- },
\end{gathered}
\end{equation}
which leads to
\begin{equation}
\begin{gathered}
    \label{Yexpr}
    Y \equiv \left< \frac{\tilde{q}^2+\tilde{j}^2}{2} \right> = \left< \frac{ F^{\prime \, 2}_{+} + F^{\prime \, 2}_{-} }{1+ F^{\prime}_{+} F^{\prime}_{-} - \textbf{X}^{\prime}_+ \cdot \textbf{X}^{\prime}_- } \right>,
\end{gathered}
\end{equation}
where we have introduced the weighed averages $\langle A \rangle \equiv \int A \tilde{\varepsilon} ds/\int  \tilde{\varepsilon} ds$, performed over all loops in the network, and $\tilde{\varepsilon}^2 = {\textbf{X}_s^{\prime \, 2}}/{(1 - \dot{\textbf{X}}^2 )}$. Expression (\ref{Yexpr}) is valid for the transonic and linear chiral models, which are the subject of study in this paper.

The relation between $Y$ and the microscopic variables, in general, depends on the particular loop configuration under consideration. However, since it can be shown that
\begin{equation}
    \label{AvePM}
    \left< \textbf{X}^{\prime}_+ \cdot \textbf{X}^{\prime}_- \right> = 0,
\end{equation}
we may safely assume that for the chiral case, on average, the amplitude of the current in the string network is given by 
\begin{equation}
\begin{gathered}
    \label{Yexpr2}
    Y \approx \left< F^{\prime \, 2}_{\pm}  \right> = \frac{\int \tilde{\varepsilon } F^{\prime \, 2}_{\pm} d s }{ \int \tilde{\varepsilon } d s }\,.
\end{gathered}
\end{equation}
This relation was previously used in Ref.~\cite{Paper1}.

\subsection{Energy loss and evolution of current-carrying loops}\label{subsec:loopevo}

The total energy of the cosmic strings may be expressed as~\cite{RybakAvgoustidisMartins} 
\begin{equation}
    \label{Energies}
    E = E_0 \left( \mathcal{F} - \left( 2 Y + K \right) \frac{d \mathcal{F}}{d K}  \right),
\end{equation}
where $E_0=\mu_0 a \int \tilde{ \varepsilon } ds$ is the core or bare energy (excluding the contribution of current) and $\mathcal{F}(K)$ is the previously introduced averaged equation of state. As demonstrated in Ref.~\cite{MPRS3}, for most physically relevant situations, we may use the linear equation of state: $\mathcal{F} = 1- K/2$, which simplifies Eq.~\eqref{Energies} to
\begin{equation}
    \label{EnergiesL}
    E = E_0 \left( 1 + Y \right), 
\end{equation}
which is then proportional to the bare energy of the string.

Let $\ell \equiv \int \tilde{\varepsilon} ds$ be the invariant loop length, which is proportional to its bare energy $E_0=\mu_0 \ell$, and $\mathcal{Y}$ be its current amplitude (while $Y$ is reserved for the network). The invariant length of the loop $\ell$ should decrease as result of the emission of gravitational and vector radiation as follows
\begin{equation}
    \label{LoopLength}
    \left. \dot{\ell} \right\rvert_{\rm rad} =  - G \mu_0 \ggr(\mathcal{Y}) - \te^2 \gem(\mathcal{Y} )\,,
\end{equation}
while the initial loop length is given by
\begin{equation}
\label{lengthIn}
    \ell(t_b)  = \alpha \xi(t_b),
\end{equation}
where  $0 < \alpha < 1$ determines the loop production length, $t_b$ represents the time of loop birth, and $\ggr(\mathcal{Y})$ and $\gem(\mathcal{Y})$ denote the efficiency of gravitational and vector emission, respectively.  The efficiency of GW emission is given by \cite{Paper1}
\begin{equation}
\label{GamGr}
  \ggr(\mathcal{Y}) = \ggr_{0} (1- \sqrt{\mathcal{Y}})^{B} ,   
\end{equation}
where $B\approx 2$ and $\ggr_0 \approx50 $.

The separation between string core and current energy described by expression \eqref{EnergiesL} implies that we should have that $\left. \dot{E}\right\rvert_{\rm rad} = \left.\dot{E}_0 \right\rvert_{\rm rad}$. This means that the total charge in the loop, in the absence of leakage mechanisms, should be conserved as loops radiate vector and gravitational radiation
\begin{equation}
    \label{FullEnergyRad}
    \left.\dot{\mathcal{Y}}\right\rvert_{\rm rad}=-\frac{\mathcal{Y}}{\ell}\left. \dot{\ell}\right\rvert_{\rm rad}=\frac{\mathcal{Y}}{\ell}\left[G \mu_0 \ggr(\mathcal{Y}) + \te^2 \gem (\mathcal{Y})\right]\,.
\end{equation}
Therefore $\mathcal{Y}$ increases as the loop shrinks.

A loss of charge, however, may occur due to various phenomena, but this is strongly dependent  on the underlying field theory model~\cite{Barr:1987xm, BARR1987146, PERKINS1991237, Gangui:1997bi, Blanco-Pillado:2002vwq, LEMPERIERE2003511, Ibe2021, Abe2023}. Here, we represent the charge loss in the same simple form we have adopted for the long string network. This choice is motivated by the study in Ref.~\cite{Ibe2021}, where it was demonstrated that the charge loss is inversely proportional to the radius curvature of the string
\begin{equation}
   \left. \dot{\mathcal{Y}} \right\rvert_{\rm leak} = - A(\mathcal{Y}) \frac{\mathcal{Y}}{\ell},
\end{equation}
where we assume that $A(\mathcal{Y)}$ has the same form as that of the network in Eq.~\eqref{Af} and assume that the critical current for loops is the same as for long strings $\mathcal{Y}_{\rm cr} = Y_{\rm cr}$ (and $ \mathcal{Y}_{\rm cr}  = {2 m^2_{\sigma}}/{3 m^2_H}$ for the Witten model).

Thus, the full system of equations describing the evolution of current-carrying loops evolution may then be written as
\begin{subequations}
\label{lYeq} 
\begin{align}
    \label{leq}
   \dot{\ell} & =  - G \mu_0 \ggr(\mathcal{Y}) - \te^2 \gem(\mathcal{Y} ), \\
   \label{Yeq} \dot{\mathcal{Y} } & = \frac{\mathcal{Y}}{\ell}\left[G \mu_0 \ggr(\mathcal{Y}) + \te^2 \gem (\mathcal{Y})-A(\mathcal{Y})\right] .
\end{align}
\end{subequations}

\begin{figure}[t]
\centering
		\includegraphics[scale=0.53]{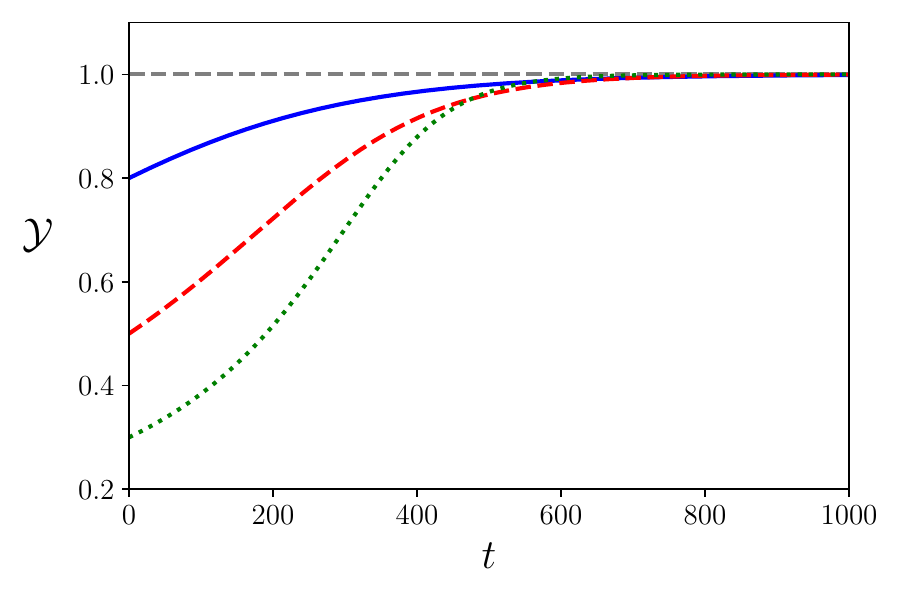}
  		\includegraphics[scale=0.53]{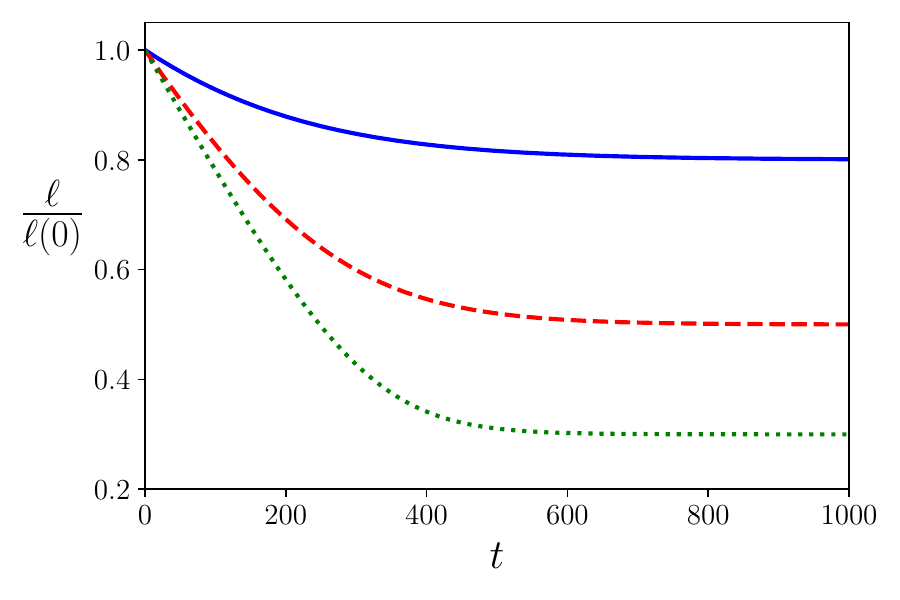}
\caption{ \label{fig:ChargeLoopEvolVorton} Evolution of superconducting loops toward vorton solutions for different values of initial current. Here we took $G \mu_0 = 10^{-10}$, $\ggr_0 = 50$, $B=2$, $\gem_0 = 9$, $D = 1$, $\tilde{e}=10^{-1}$, $A_{\rm const} = 0$. The top panel represents the evolution of charge $\mathcal{Y}$, while the bottom panel displays the evolution of the invariant length of the loop.}
\end{figure}

These equations admit attractive steady-state solutions of the form
\begin{equation}
    \dot{\ell}=\mbox{constant}\quad\mbox{and}\quad \dot{\mathcal{Y}}=0\,.
\end{equation}
The value of the scaling constants depends on both the initial conditions and the parameters of the model ($\mu_0, e, \mathcal{Y}_{\rm cr}$) - or, in other words, on the relative importance of the different physical processes acting on the loop.

If charge leakage is very efficient, current is quickly dissipated and  $\mathcal{Y} \rightarrow 0$. In this limit, the emission of electromagnetic radiation is suppressed (cf. Eq.~\eqref{Gfun}) and $\ggr \to \ggr_0$. From this point on, these loops behave as Nambu-Goto loops. On the opposite limit, in the absence of leakage mechanisms (or if leakage is extremely inefficient), total charge conservation implies that $\mathcal{Y}\propto 1/\ell$ and that current grows monotonically until it eventually reaches $\mathcal{Y} \rightarrow 1$. In this case, we have that $\ggr,\gem \to 0$ and the emission of radiation ceases. From this point on, the length of the loop remains constant and equal to $\ell_f = \ell(0) \mathcal{Y}(0)$ and the loop turns into a vorton. We display the evolution of such loops in Fig.~\ref{fig:ChargeLoopEvolVorton}.

\begin{figure}[t]
\centering
		\includegraphics[scale=0.53]{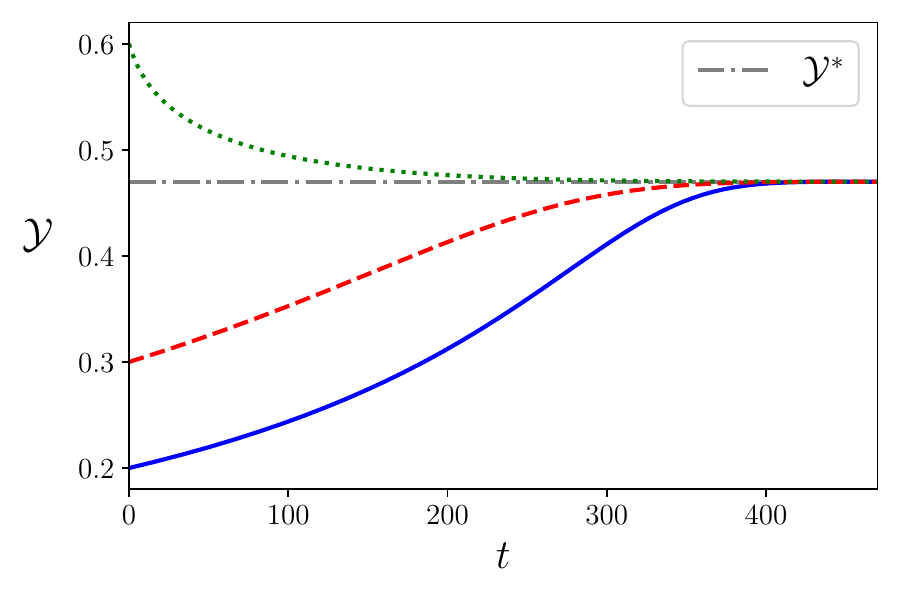}
  		\includegraphics[scale=0.53]{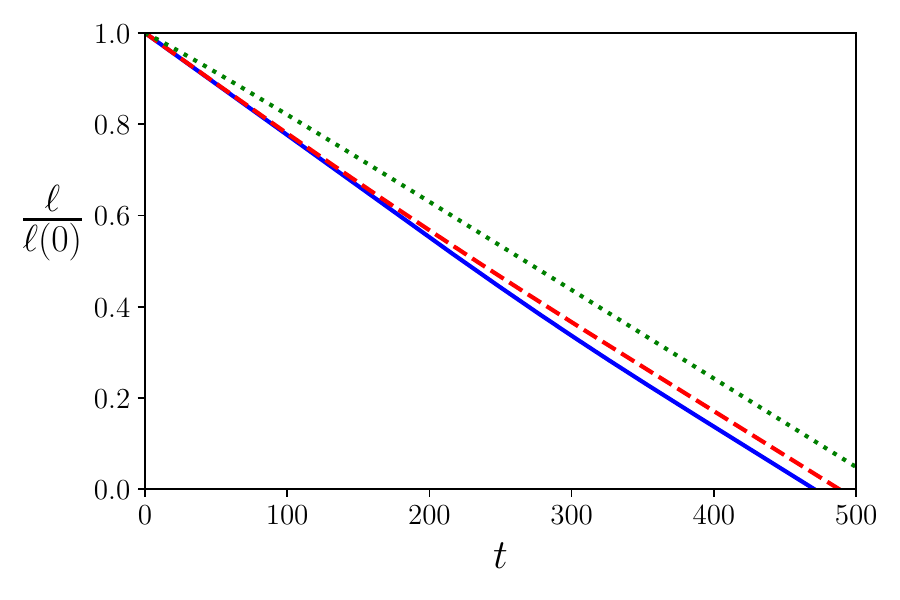}
\caption{ \label{fig:ChargeLoopEvol} Evolution of current-carrying loops toward a scaling solution with constant current, for different values of initial current. Here, we took $G \mu_0 = 10^{-10}$, $\ggr_0 = 50$, $B=2$, $\gem_0 = 9$, $D = 1$, $\tilde{e}=10^{-1}$, $\mathcal{Y}_{\rm cr} =0.7$, $A_{\rm const} = 10^{-3}$. The top panel represents the evolution of charge $\mathcal{Y}$, while the bottom panel displays the evolution of the invariant length of the loop.}
\end{figure}

Moreover, Eqs.~\eqref{lYeq} admit a scaling solution with a constant non-vanishing current $\mathcal{Y}^*$ that satisfies the algebraic equation
\begin{equation}
    \label{YsolGen}
   G \mu_0 \ggr(\mathcal{Y}^*) + \te^2 \gem (\mathcal{Y}^*) = A(\mathcal{Y}^*) .
\end{equation}
We display, in Fig.~\ref{fig:ChargeLoopEvol}, examples of loops evolving toward this scaling solution. We may see from Eq.~\eqref{YsolGen} that the scaling value $\mathcal{Y}^*$ is entirely determined by the charge leakage function $A(\mathcal{Y})$, as the other parameters are fixed. Therefore, we anticipate that a loop produced by a cosmic string network with an initial current amplitude $\mathcal{Y}(0)$ should maintain this value, at least during the initial stages of loop decay. There is no reason for the charge leakage function $A(\mathcal{Y})$ to undergo abrupt changes at the moment of loop formation. Only when the string loop experiences significant changes in its length it would be possible for the leakage function to decrease to zero --- leading to the formation of a vorton --- or for the leakage to become more efficient, resulting in fast charge escape. All these scenarios require that the leakage function $A(\mathcal{Y})$ exhibits explicit length dependence, which becomes evident at a particular scale. In this study, we will restrict ourselves to cases in which the initial value of the charge amplitude remains constant throughout the evolution of the loop, i.e., $\mathcal{Y} \approx \mathcal{Y}^*$.

\subsection{Period of oscillation for superconducting loops } \label{osc}

The presence of a current introduces inertia to the motion of cosmic string loops, reducing their velocity and increasing the duration of one oscillation. This effect can be explicitly observed in the Burden and kinky loop solutions if one fixes the bare length of the loop, which is given by $G_{\pm} \ell$ in the symmetric case and $G_{+} \ell$ (or $G_- \ell$) in  the chiral case for Eqs.~\eqref{ABmodesLoops1} and \eqref{CuspLessLoop}, respectively. The period of oscillation of the loops increases as the current grows and tends to infinity whenever the current saturates (i.e., $T \rightarrow \infty$ as $G_{\pm} \rightarrow 0$).

In the case of kinky loops, defined by Eq.~\eqref{CuspLessLoop}, we can maintain the length of the loop fixed by redefining the parametrization: $\tilde{\sigma}_{\pm}={\sigma_{\pm}}/{G_{\pm}}$. Thus, the new periodicity of $\tilde{\sigma}_{\pm} = \tilde{\sigma}_{\pm} + {2 \pi}/{G_{\pm}}$ leads to a new periodicity in $\tilde{\tau} = \tilde{\tau}+2 \pi {(G_+ + G_-)}/{(G_+ G_-)}$. One may then conclude that the period of oscillation $T$ of kinky loops is related to the invariant length of the loop through the expression
\begin{equation}
\label{PeriodOscillationCuspLess}
    T = \ell \frac{G_- + G_+}{4 G_- G_+},     
\end{equation}
while the frequency of the $j$-th mode is given by
\begin{equation}
    \label{FreqOscillationCuspLess}
    f = \frac{4 j G_- G_+}{ \ell_j ( G_- + G_+ )}.    
\end{equation}
For a symmetrical current, with $G_+ = G_-$, Eq. \eqref{PeriodOscillationCuspLess} is valid for Burden loops as well and it also correctly reproduces the limit $G_{\pm} \rightarrow 0$. The general relationship between the period and string length for current-carrying string loops can vary depending on their shape. However, the condition $T \rightarrow \infty$ in the limit ${G_{\pm} \rightarrow 0}$ holds universally. Eq.~\eqref{PeriodOscillationCuspLess} reproduces these limits correctly, and since we do not expect significant deviations from this relationship for other loop shapes, we will use Eq.~\eqref{PeriodOscillationCuspLess} in our work. Nevertheless, further studies, particularly numerical simulations, could provide deeper insights into the evolution of current-carrying cosmic string loops \cite{Martin:2000ca, Cordero-Cid:2002hmv, Battye:2021sji, Battye:2021kbd, Battye:2022mxi}.

\section{SGWB generated by superconducting chiral cosmic strings}

For simplicity, throughout this section, we will assume that the current on the loops remains constant, with $0\le\yl<1$, and coincides with the current of the long string network at the moment of creation. In other words, we assume that loops are in the scaling regime described in Sec.~\ref{subsec:loopevo}. Notice that this may be precisely the situation in which specific signatures of current --- if any --- would arise in the SGWB. As a matter of fact, if leakage is very efficient and loops lose current rapidly, their SGWB should be similar to that generated by standard strings. Moreover, if loops quickly evolve toward a vorton-like solution, their GW emission halts and, consequently, only a fraction of their energy will be converted into GWs (see Refs.~\cite{Auclair:2020wse, Auclair:2022ylu} for more details). Naively, this would result in a weaker SGWB~\footnote{However, since $\ggr\to 0$ as loops reach the $\mathcal{Y}\to 1$ limit in which the amplitude of the SGWB may be enhanced, the situation may be more complex. We leave the case of vortons for future work.}.

We will also assume that, on average, the cosmic string network does not have any bias between time-like and space-like current loss mechanisms, which leads to $K \approx 0$ \cite{MPRS3}. Thus, we will be considering a chiral cosmic string network that produces chiral loops with $G_+ = 0$ (or $G_- = 0$). We leave more generic cases for further studies.

Although the SGWB generated by current-carrying string loops with constant current was studied preliminarily in Ref.~\cite{Paper1}, this study did not include the emission of vector radiation and did not take the effect of current on the frequency of oscillation of loops fully into account. Here, we will include both these effects and discuss in detail their impact on the SGWB. Throughout this section, we will assume that gravitational and vector radiation are characterized, respectively, by $\gem_0=9$ and $D=1$, and $\ggr_0=50$ and $B=2$.

\subsection{Formalism for loops with constant current} \label{Formalism}

The amplitude of the SGWB is often characterized in terms of the spectral energy density of gravitational radiation, in units of the critical density of the universe $\rho_c$,
\begin{equation}
\Omega_{\rm gw}(f)=\frac{1}{\rho_c}\frac{d\rho_{\rm gw}}{d \log{f}}\,,
\end{equation}
where $f$ is the frequency measured by an observer at the present time and $\rho_c=3H_0^2/(8\pi G)$. At any given frequency $f$, the SGWB comprises contributions from all loops that have emitted GWs throughout cosmic history, which arrive at the present time to an observer with frequency $f$. This means that $\Omega_{\rm gw}(f)$ has contributions from all the loops with a length that satisfies
\begin{equation}
\label{LoopLength2}
\ell_j(t)=\frac{4 j \sqrt{1-\mathcal{Y}} }{ f ( 1 + \sqrt{1-\mathcal{Y}} )} \frac{a(t)}{a_0}\equiv \frac{2j}{f}\frac{a(t)}{a_0}S(\yl)\,,
\end{equation}
where we used Eqs.~\eqref{Yexpr2} and \eqref{FreqOscillationCuspLess}. 

We then have that (see e.g.~\cite{Vilenkin:2000jqa})
\begin{equation}
\label{FullOmega}
    \Omega_{\rm gw}(f)=\frac{16\pi}{3}\left(\frac{G\mu_0}{H_0}\right)^2 \sum^{n^*}_{j=1}  \Omega_{\rm gw}^j(f)\,,
\end{equation}
where
\begin{equation}
\label{OmegaJ}
    \ogw^j (f) = \frac{j}{f}\int_{t_f}^{t_0}  \left(\frac{a(t')}{a_0}\right)^5 \, P^{\rm gw}_j(\yl) n(\ell_j(t'),t') dt'
\end{equation}
describes the contribution of the $j$-th harmonic mode of emission to the SGWB. Here, $t_f$ is the time in which significant emission by cosmic string loops begins, which we assume roughly coincides with the end of the friction era~\footnote{Note however that it was recently shown that, for strings without current, in some situations, loops created during friction may contribute significantly to the SGWB as well~\cite{Mukovnikov:2024zed}.}, and $n_*$ is the number of harmonic modes taken into consideration in the computation of the SGWB. Moreover, $n(\ell,t)d\ell$ is the number density of loops with lengths between $\ell$ and $\ell+d\ell$ that exist at a time $t$ and this is the key ingredient in the computation of the SGWB. Note that, when one considers current, the normalization of the spectrum of emission $P^{\rm gw}_j$ (or in other words, the GW emission efficiency) depends on the current of the cosmic string loops:
\begin{equation}
 \label{PowerRadGr}
   P^{\rm gw}_j(\mathcal{Y})\equiv \Gamma^{\rm gw}_0  j^{-q} \text{e}^{-j g(\mathcal{Y})}    \frac{(1-\mathcal{Y}^{1/2})^B}{\mathcal{E}(\mathcal{Y})},
\end{equation}
with~\cite{Paper1}
\begin{equation}
\label{gCK}
g(\mathcal{Y}) \approx \begin{cases} 8 \left( 1 - \left( 1-\mathcal{Y} \right)^{1/4} \right)^{2} , & \quad q=4/3  \\ & \text{(quasi-cusps)} \\
0 \;, \qquad & \quad q=5/3 \\
& \quad \, \text{(kinks)}
\end{cases}
\end{equation}
and $\mathcal{E}(\mathcal{Y}) = \sum_{j=1}^{n^*} j^{-q} \text{e}^{-j g(\mathcal{Y})}$. Here, since for both loops with kinks and quasi-cusps, $P^{\rm gw}_j$ decreases with increasing harmonic mode, in practice one may consider a finite number of harmonic modes (instead of considering $n^*\to + \infty$)\cite{Sanidas:2012ee}. We have verified that taking $n^* \approx 10^4$ is sufficient in both cases. 

The number density of loops may be found using the semi-analytical method introduced in~\cite{SousaAvelino}:
\begin{equation}
    n(\ell,t)\frac{d\ell_b}{dt_b}=\left.\frac{n_c}{dt}\right|_{t=tb}\left(\frac{a(t_b)}{a(t)}\right)^3
\end{equation}
where $\ell_b$ is the length of loops at the time of creation $t_b$ and $n_c$ is the number density of loops created. If one assumes that all loops are created with the same length and that this is determined by the characteristic length of the network --- i.e. that $\ell_b=\alpha \xi(t_b)$, with constant $0<\alpha<1$ --- we should have that~\cite{SousaAvelino}
\begin{equation}
    \frac{dn_c}{dt}=\frac{\mathcal{F}_{\rm fuzz}}{\sqrt{2}}\frac{c v}{\alpha \xi^4}\,,
\end{equation}
where the factor of $1/\sqrt{2}$ was introduced to account for the red-shifting of the peculiar velocities of loops~\cite{Vilenkin:2000jqa} and we have introduced the \textit{fuzziness} parameter $\mathcal{F}_{\rm fuzz}$ as a way to account for potential uncertainties in the normalization. Such a parameter may describe, for instance, situations in which not all loops are created with the same length and velocity~\cite{Blanco-PilladoOlum,Sousa:2020sxs,Blanco-Pillado:2024aca} or in which only a fraction of loops contributes to the SGWB~\cite{Hindmarsh:2022awe}. Note that $\alpha$ and $\mathcal{F}_{\rm fuzz}$ are currently unknown for realistic superconducting string networks, as it was not yet possible to study loop production with numerical simulations for this type of string. Here, we will use the values inferred from Nambu-Goto simulations~\cite{Blanco-PilladoOlum}--- $\alpha\approx 0.34$ and $\mathcal{F}_{\rm fuzz}=0.1$ --- as fiducial values for our study, but we will also discuss the impact of changing the value of $\alpha$ on the results~\footnote{The impact of the fuzziness parameter is trivial, as the amplitude of the spectrum scales proportionality to $\mathcal{F}_{\rm fuzz}$.}. As pointed out in~\cite{Paper1}, since the large scale dynamics of the cosmic string network is significantly affected by current (cf. Sec.~\ref{CVOS model}), so should $n_c$ be --- particularly in the $\yl\to 1$ limit, in which the network becomes increasingly dense and the number of loops produced steeply grows. However, now that we also include vector radiation as a decay mechanism, loops evolve differently, which affects the normalization of $n(\ell,t)$ as well, through the term
\begin{equation}
\begin{gathered}
    \frac{d\ell_b}{dt_b}=\alpha \frac{d\xi_b}{dt_b}+G\mu_0 \ggr(\yl_b)+\te^2 \gem (\yl_b) \equiv \\
    \equiv \alpha \frac{d\xi_b}{dt_b}+P(\yl_b)\,,
\end{gathered}
\end{equation}
where we have introduced the total power emitted by the loop $P(\yl)$, including the emission of both gravitational and vector radiation in all modes of emission, 
\begin{equation}
    \frac{d\xi_b}{dt_b}= \epsilon_b \left( 1 + \frac{v_b^2}{1+Y_b} \right) + \frac{c v_b}{2} + Y_b \frac{v_b k(v_b)}{1+Y_b}.
\end{equation}
and the subscript `$b$' is used to indicate that the corresponding quantities are evaluated at the time of birth of loops. Notice that, since loops are assumed to be created with the same average current as that of the long string network, $\yl_b=Y_b$. Once the large scale evolution of $\xi$ (or $\epsilon$), $v$, and $Y$ is fully characterized, the computation of the SGWB at any given frequency $f$ then simply involves determining the time of birth of loops that have a length $\ell_j$ at the time $t$, by solving numerically the equation
\begin{equation}
    \frac{2j}{f}\frac{a(t)}{a_0}S(Y_b)=\alpha \xi_b + P(Y_b)(t_b-t)\,,
\end{equation}
where we have used the fact that current remains constant during the decay of the loops.

\subsection{Basic spectral shape}

Before we venture into the full computation of the SGWB, we will start by discussing the basic shape of the spectrum and by identifying its main features. The SGWB generated by cosmic string loops generally receives contributions from three distinct loop populations: loops created and decaying in the radiation era (radiation-era loops), loops formed in the radiation era that survive the radiation-matter transition and decay in the matter era, and loops created after this transition (matter-era loops) \cite{Sousa:2020sxs}. Since, as discussed in Sec.~\ref{CVOS model}, current affects the evolution of cosmic strings essentially during the radiation era\footnote{Except perhaps for very high values of current ($Y\sim 1$), in which the dissipation of current may be slower as the rate of expansion becomes faster.}, we should expect current to affect mostly the first two populations of loops. Also, during the radiation era, a network of current-carrying strings is, as we have seen, expected to evolve in a linear scaling regime with constant current, which means that all loops created during this stage should decay at the same rate. Current-carrying string networks should also then generate a scaling population of loops during the radiation era --- whose number density may be expressed in the form $n(\ell,t)=t^{-4}n(\ell/t)$~\cite{LISA} --- and the general shape of the contribution of these two loop populations to the SGWB should be hence similar to that of standard strings (however with different amplitude).

The SGWB spectrum generated by current-carrying networks then should also have a plateau at high frequencies, generated by the first population of loops (those that decay in the radiation era). Assuming then that the network is, during the radiation era, in a linear scaling regime characterized by constant $\zeta\equiv\xi/t=\epsilon/(Ht)$, $v$ and $Y$, we find, following the approach in~\cite{Sousa:2020sxs}, that the amplitude of the plateau is given by\footnote{Strictly speaking, this expression is valid for $Y_\rd \ne 1$ as, in its derivation, it is assumed that cosmic string loops decay. When $Y_\rd=1$ there is no emission of gravitational and vector radiation by the loops and, therefore, $\ogw=0$.}
\begin{equation}
\begin{gathered}
\label{eq:plateau}
    \Omega_{\rm gw}^{\rm plateau} =\frac{128 \pi}{9}  A_\rd \Omega_{\rm r} \frac{\ggr(Y_\rd)(G\mu_0)^2}{\alpha \zeta_\rd S(Y_\rd)} \times \\ 
   \times \left[\left(\frac{\alpha \zeta_\rd}{P(Y_\rd)}+1\right)^{3/2}-1\right]\,,
\end{gathered}
\end{equation}
where we have defined $A_\rd=c\mathcal{F}_{\rm fuzz}v_\rd/(\sqrt{2}\alpha \zeta_\rd^3)$ and the subscript `$\rd$' is used to indicate that the corresponding quantity is evaluated during the radiation era. This expression allows us to precisely highlight the impact of the two effects we introduced on the amplitude of the spectrum.

In particular, the change in the frequency of oscillation of cosmic string loops introduced by the inertia of the current carriers leads, not only to a shift in the frequency of the spectrum --- as one would naively expect ---, but also to an enhancement of the amplitude of the spectrum by a factor of $1/S(Y_\rd)$ (notice that $0<S(Y_\rd)<1$). This may be explained by the fact that loops that contribute to a given frequency at a time $t$ have to be created later in cosmic history than in the absence of current (and the higher the current the later they have to be created) and thus their GWs reach an observer at $t_0$ less diluted by expansion.

\begin{figure}[t]
\begin{center}
\includegraphics[width=0.98\linewidth]{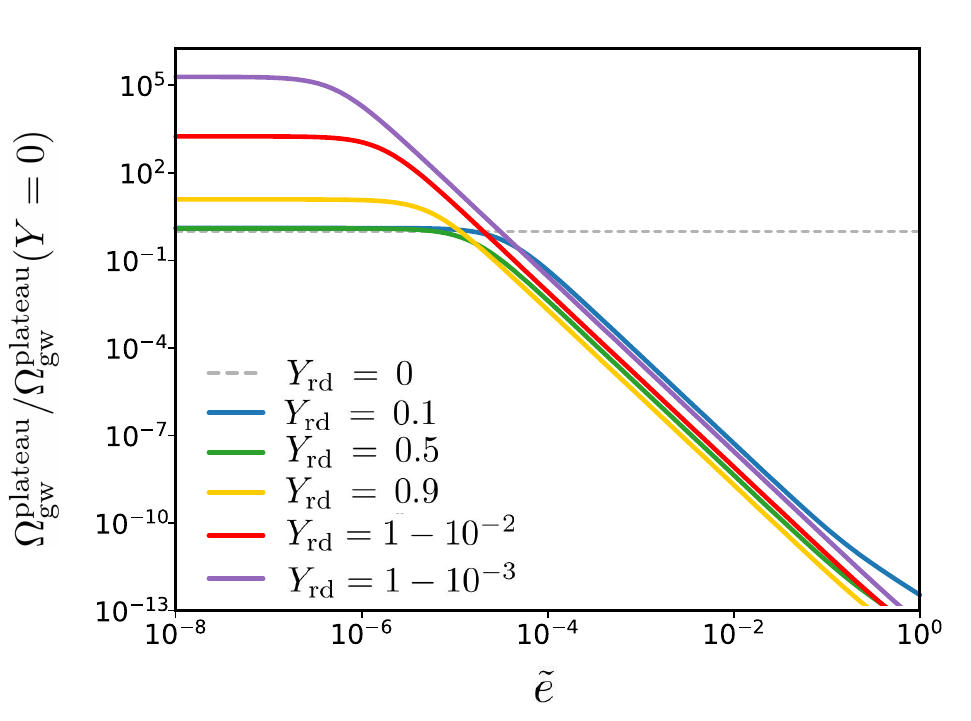}
\caption{\label{Fig:impactq}Dependence of the amplitude of the radiation era plateau on the value of the charge of the current carriers for different values of current, normalized to the amplitude of the plateau of standard strings. Here we took $G\mu_0=10^{-10}$, $\alpha=0.34$, and $\mathcal{F}_{\rm fuzz}=0.1$.}

\end{center}
\end{figure}

As to the impact of the emission of vector radiation, if $\alpha\zeta_\rd\gg P(Y_\rd)$, it is straightforward to see from Eq.~\eqref{eq:plateau}
that the amplitude of the plateau of the spectra should be independent of $\te$ and $\gem_0$ if $\ggr(Y_\rd) G\mu_0 \gg \te^2\gem (Y_\rd)$. This corresponds to the scenario discussed in~\cite{Paper1} (except for the impact of the shift of frequency that was not taken into consideration therein), as the emission of vector radiation has a negligible impact in this case. However, when $\te^2\gem (Y_\rd)$ exceeds $G\mu_0 \ggr(Y_\rd)$, the amplitude of the spectrum decreases steeply with increasing charge as $\Omega_{\rm gw}^{\rm plateau}  \propto \te^{-3}$. This is illustrated in Fig.~\ref{Fig:impactq}, where we plot the ratio between the amplitude of the radiation era plateau for current-carrying strings and Nambu-Goto strings as a function of $\tilde{e}$ for different values of $Y_\rd$. This plot clearly demonstrates these two regimes, but deviations from the $\Omega_{\rm gw}^{\rm plateau}  \propto \te^{-3}$ trend appear at large $\tilde{e}$. These deviations arise because, in this limit, the condition $\alpha \zeta_\rd \gg  P(Y_\rd)$ is no longer satisfied. Additionally, for values of $\tilde{e}$ in which vector radiation dominates the decay of loops, the amplitude of the plateau is less sensitive to $Y_\rd$ (as the steep increase in the $Y_\rd \to 1$ limit is slower). Moreover, the results suggest that for strings carrying an electromagnetic current (with $\tilde{e} \sim 10^{-2}-1$), the SGWB amplitude is significantly suppressed, except in the very high-current limit.

Similarly, the general shape of the contribution generated by the population of radiation-era loops that survive into the matter era is also not significantly affected by current, so it gives rise to peak in the low frequency range of the spectrum as in the case of currentless strings. However, this contribution will exhibit a similar shift in frequency and a similar dependence on current as the radiation-era plateau.

\begin{figure}[t]
\begin{center}
\includegraphics[width=0.98\linewidth]{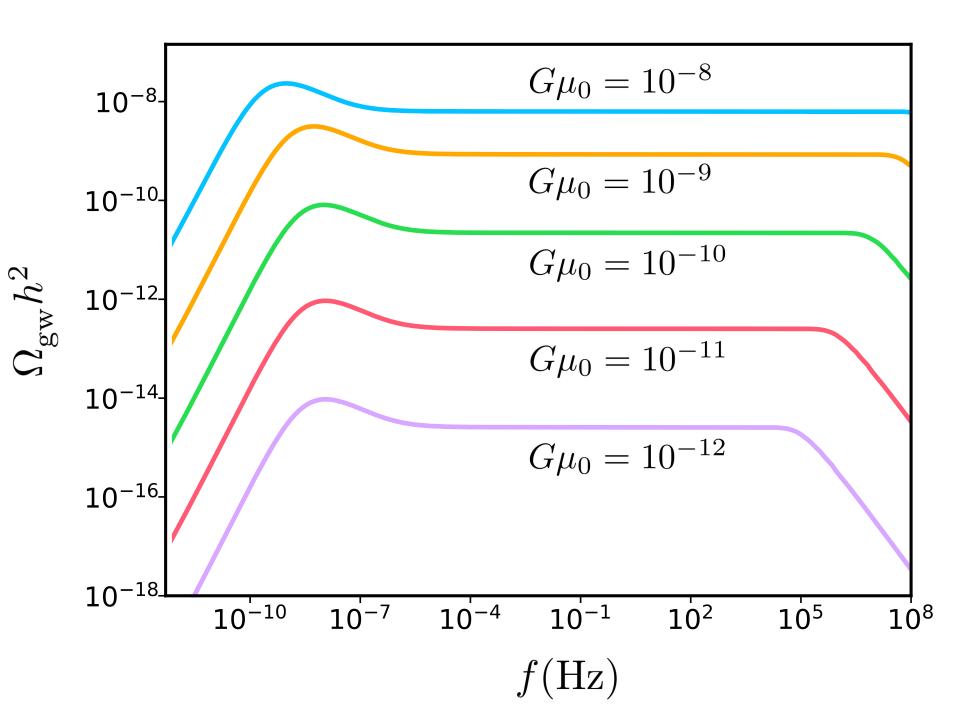}
\caption{\label{Fig:Gmus}Impact of $G\mu_0$ on the SGWB generated by current-carrying strings. Solid thin lines represent the spectrum computed using the one parameter approximation for the CVOS described in Sec.~\ref{subsec:cvosapp}, while thick transparent lines represent the spectra computed using the full CVOS model. Here we took $\alpha=0.34$, $\mathcal{F}_{\rm fuzz}=0.1$, $\tilde{e}=10^{-4}$ and $Y_\rd=0.7$ and include only the fundamental mode of emission (i.e. $n_*=1$).}

\end{center}
\end{figure}

Notice that now, when one varies the tension of cosmic strings, the relative importance of the loop decay mechanisms is altered. As a matter of fact, as one decreases $G\mu_0$ --- even if $\te$ is small --- eventually vector radiation will be the dominant decay channel for loops (or, in other words, eventually $\te^2 \gem\gg\ggr G\mu_0$). Once this regime is reached, as may be seen from Eq.~\eqref{eq:plateau}, the amplitude of the plateau decreases as $G\mu_0^2$ with decreasing tension --- significantly faster than what is seen when the emission of gravitational radiation dominates ($\Omega_{\rm gw}^{\rm plateau} \propto (G\mu_0)^{1/2}$). Moreover, in the regime in which vector radiation dominates, the peak of the spectrum no longer shifts toward higher frequencies since now the lifetime of the loops (which is the primary factor that determines the peak frequency of the radiation-era contribution) is determined by $\te$ and independent of $G\mu_0$. The transition between these two regimes may clearly be seen in Fig.~\ref{Fig:Gmus}, where we plot the SGWB for fixed $Y_\rd$ and $\tilde{e}$ with different values of $G\mu_0$. Notice that, in this plot (and in the rest of the plots in this section), we assume that all gravitational radiation is emitted in the fundamental mode (i.e $n_*=1$). This is justified by the fact that the impact of the emission of vector radiation on the SGWB is largely independent of the spectrum of emission of gravitational radiation. The inclusion of higher harmonics causes, as we shall see in the next section, a shift of power toward higher frequencies and, therefore, slightly affects the shape of the peak of the spectrum --- causing a small decrease in its height and its broadening (see e.g.~\cite{Sanidas:2012ee,SousaAvelino}) --- but does not affect the amplitude of the plateau. As a matter of fact, overall, the amplitude of the spectrum is dominantly determined by the total vector and gravitational radiation emission efficiencies, $\gem$ and $\ggr$, which do not depend on the spectrum of emission.

\begin{figure}[t]
\begin{center}
\includegraphics[width=0.98\linewidth]{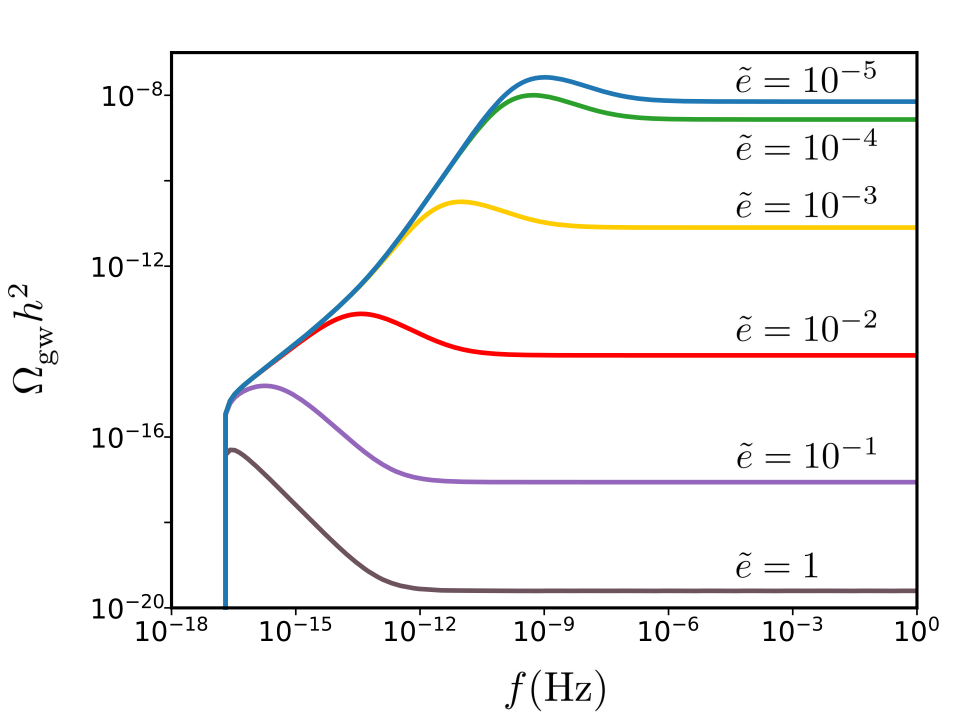}
\caption{\label{Fig:qs}Impact of $\tilde{e}$ on the SGWB  generated by current-carrying strings. Solid lines represent the fundamental mode of emission of the SGWB computed using the full CVOS model. Here we took $G\mu_0=10^{-8}$, $\alpha=0.34$, $\mathcal{F}_{\rm fuzz}=0.1$, and $Y_\rd=0.7$ and include only the fundamental mode of emission (i.e. $n_*=1$).}
\end{center}
\end{figure}


Up to now, we have only considered situations in which the dominant contribution to the spectrum comes from radiation-era loops. However, the loops that are created and decay after the radiation-matter transition also generate a peaked contribution in a similar frequency range and the final shape and height of the peak of the spectrum depends on the contributions of both loop populations. For Nambu-Goto strings, which carry no current, the contribution of matter era loops is generally negligible for values of $G\mu_0$ compatible with current observational constraints. This is not, however, necessarily the case for current-carrying strings: charge is rapidly dissipated after the radiation-matter transition and therefore the matter era contribution should not be as significantly affected by current and the emission of vector radiation. In some situations, this results in changes to the shape of the peak of the spectrum.

Significant signatures appear, for instance, for large enough $\te$, as in this case the amplitude of the spectra generated by radiation-era loops is significantly suppressed. In Fig.~\ref{Fig:qs}, we plot the SGWB generated by current-carrying strings with $Y_\rd=0.7$ and $G\mu_0=10^{-8}$ for different values of $\te$. Therein, one may see that as $\te$ decreases and we enter the regime in which vector radiation dominates the decay of loops, the height of the peak relative to the plateau increases since the contribution of matter era loops becomes increasingly relevant. When $\te=10^{-1}$, the amplitude of these two contributions is, in fact, very similar, which leads to significant change to the shape of the peak of the spectrum. Notice that when $\te=1$, we have that $\te^2\gem \gg \alpha \zeta$, in general, and as a result loops decay effectively immediately after production by emitting vector radiation. In this regime the peak of the spectrum becomes sharper and the low-frequency cutoff of the spectrum becomes very pronounced, and its shape is no longer altered by increasing $\te$ even further. In this small loop regime (small in the sense that loop length is significantly smaller than the vector radiation scale), the shape of the SGWB is very similar to what one would have in the absence of currents when loops are small compared to the gravitational backreaction scale ($\ggr_0 G\mu_0 \gg \alpha \zeta $), in which loops should decay immediately after formation as well~\cite{Sousa:2014gka}. The inclusion of vector radiation then implies that the threshold for entering this regime is not only dependent on tension, but also on the charge of current carriers.

\begin{figure}[t]
\begin{center}
\includegraphics[width=0.98\linewidth]{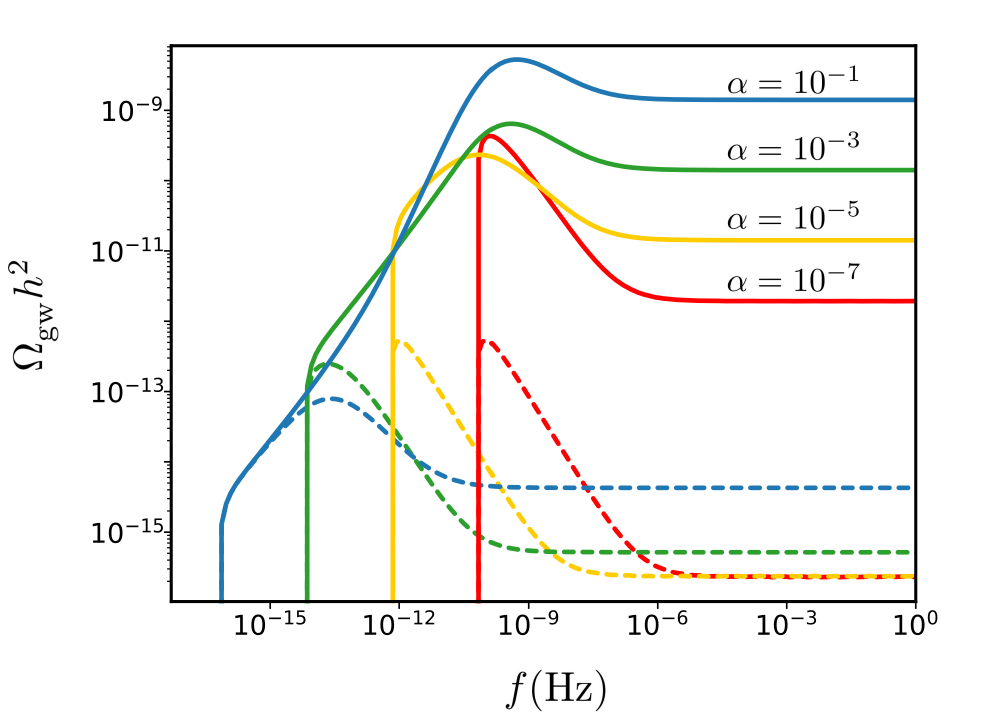}
\caption{\label{Fig:alphass}Impact of $\alpha$ on the SGWB generated by current-carrying strings. Solid lines represent the fundamental mode of emission of the SGWB computed using the full CVOS model for $\tilde{e}=10^{-4}$, while dashed lines represent the spectra for $\te=10^{-2}$. Here we took $G\mu_0=10^{-8}$, $\mathcal{F}_{\rm fuzz}=0.1$, and $Y_\rd=0.7$  and include only the fundamental mode of emission (i.e. $n_*=1$).}
\end{center}
\end{figure}

Another aspect that is known to affect the relative importance of the radiation and matter peaks is the size of loops~\cite{SousaAvelino,Sousa:2020sxs} --- which for superconducting strings, as we have already mentioned, is currently unknown --- and, in fact, a variation of $\te$ may, in a way, be regarded as a decrease of the length of the loops that is converted into gravitational radiation. So, the impact of changing $\te$ in the shape of the peak of the spectrum is, in a sense, similar to the impact of varying $\alpha$ (but, as we shall see, the impact on the overall amplitude of the spectrum is quite different). Notice, however, that, if vector radiation dominates, the threshold for entering the small-loop regime is determined by $\te$ and this may be actually reached for values of $\alpha$ that are much larger than the gravitational backreaction scale. To illustrate this, we display in Fig.~\ref{Fig:alphass}, the SGWB generated by loops with different loop sizes and two distinct values of $\te$. Therein, one may see that, indeed as one decreases the length of loops and the matter-era contribution becomes more important, the shape of the spectra changes in a similar way as we have seen when we increased $\te$: the height of the peak relative to the background increases and the low-frequency cutoff of the spectrum becomes much sharper. The decrease in the amplitude of the spectrum is, of course much slower, as it roughly scales with $\alpha^{1/2}$ in the large loop regime (as we have for currentless strings~\cite{Sousa:2020sxs}), while when we vary $\te$ it scales as $\te^{-3}$. The progressive changes to the shape of the spectrum are also very similar to what we have in the absence of current, but now this is accompanied by a shift of the peak toward lower frequencies (explained by the fact that the current shifts the radiation contributions toward higher frequencies). Note, however, that these changes to the shape of the spectra emerge for larger values of $\alpha$ when $\te$ is larger and, in fact, the loops enter a small-loop regime for larger values of $\alpha$ as well.

\subsection{The full spectrum}

\begin{figure}[t]
\centering
		\includegraphics[scale=0.5]{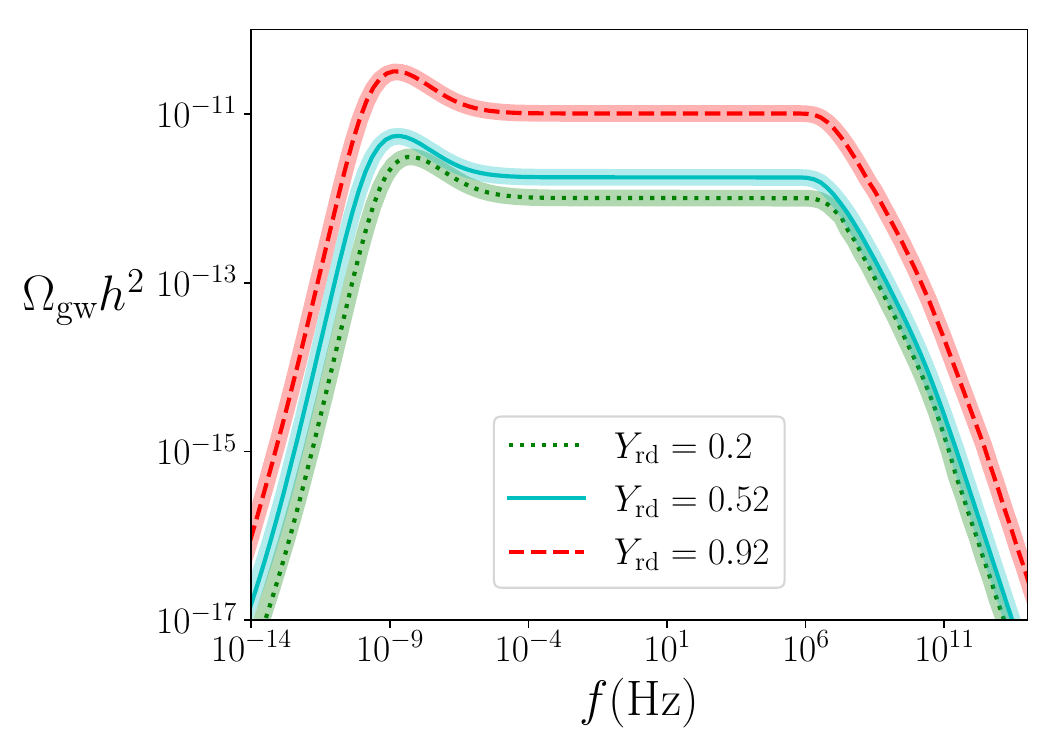}
\caption{ \label{fig:YapprComparison} The SGWB generated by loops with kinks for models with a leakage function given by Eq.~\eqref{Af} alongside that obtained using the approximate description for
$Y(x)$ defined in Eq.~\eqref{Yphen}. Here, we took $G \mu_0 = 10^{-10}$, $\tilde{e}^2 = 10^{-4}$, $\alpha=0.34$, and $\mathcal{F}_{\text fuzz}=0.1$. The values of currents and line styles correspond to those in the left panel of Fig.~\ref{fig:CVOS2}.}
\end{figure}

\begin{figure*}[t]
\begin{center}
\includegraphics[scale=0.38]{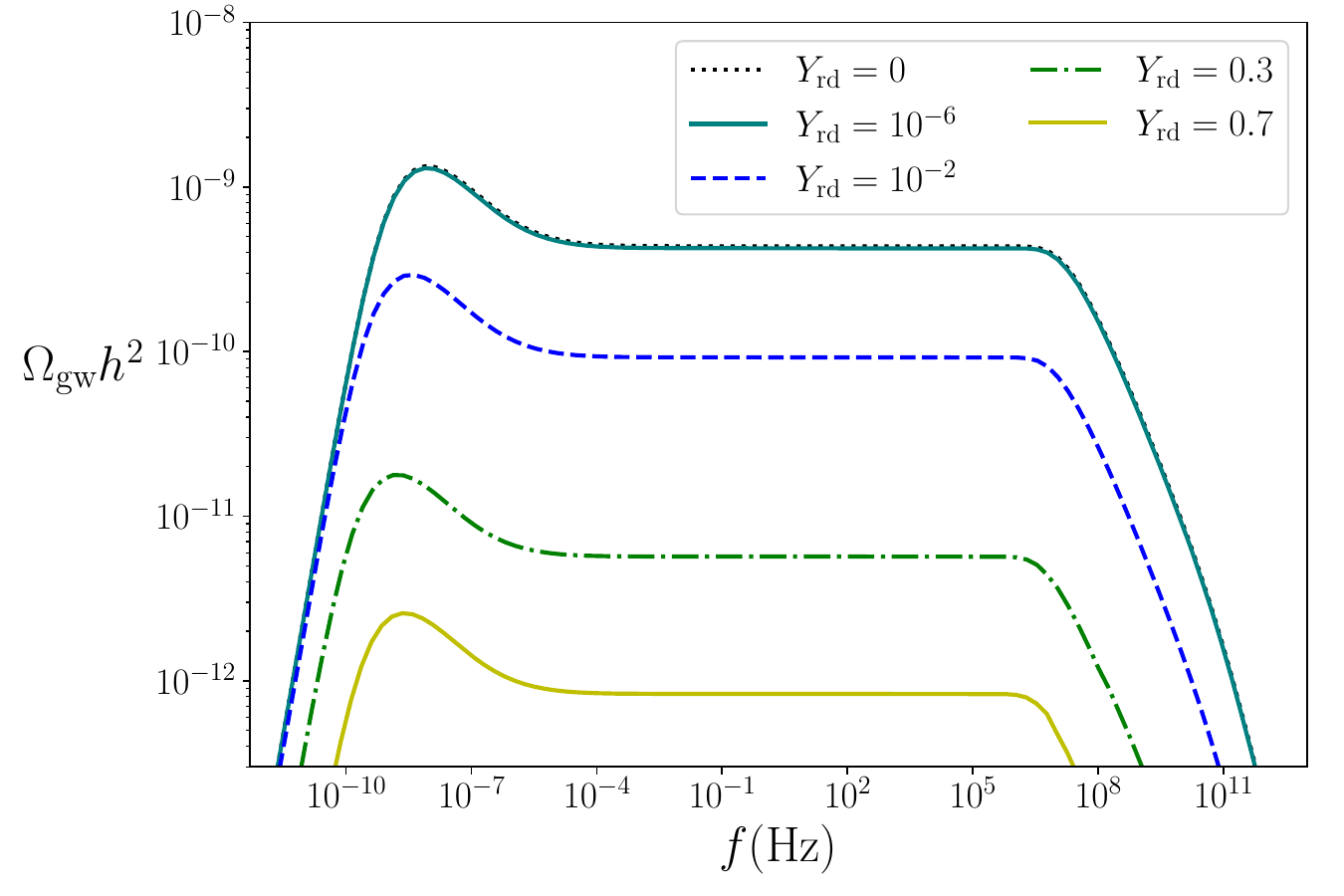}
\includegraphics[scale=0.38]{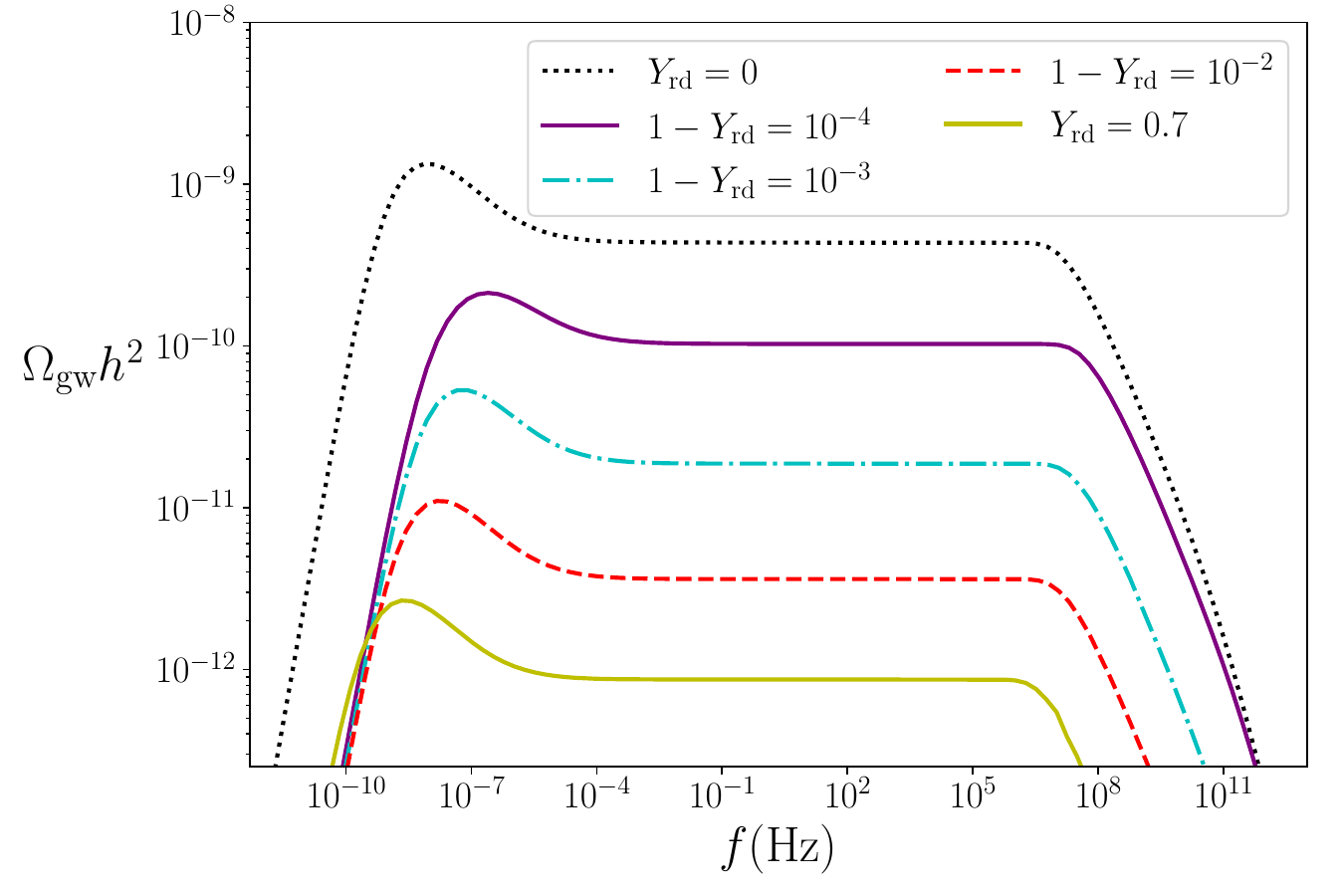}
\caption{\label{Fig:SGWB_kinks} 
The SGWB generated by current-carrying cosmic string networks for different values of $Y_\rd$, for the case in which kinks dominate the GW emission of loops. The left panel spans the range $0 < Y_{\rm rd} \leq 0.7$, while the right panel represents $0.7 \leq Y_{\rm rd} < 1$. The dotted black line corresponds to the SGWB generated by standard Nambu-Goto string network. All plots are derived for fixed parameters: $G \mu_0 = 10^{-10}$, $\alpha=0.34$, $\mathcal{F}_{\rm fuzz} = 0.1$, $\tilde{e} = 10^{-4}$.}
\end{center}
\end{figure*}

\begin{figure*}[t]
\begin{center}
\includegraphics[scale=0.38]{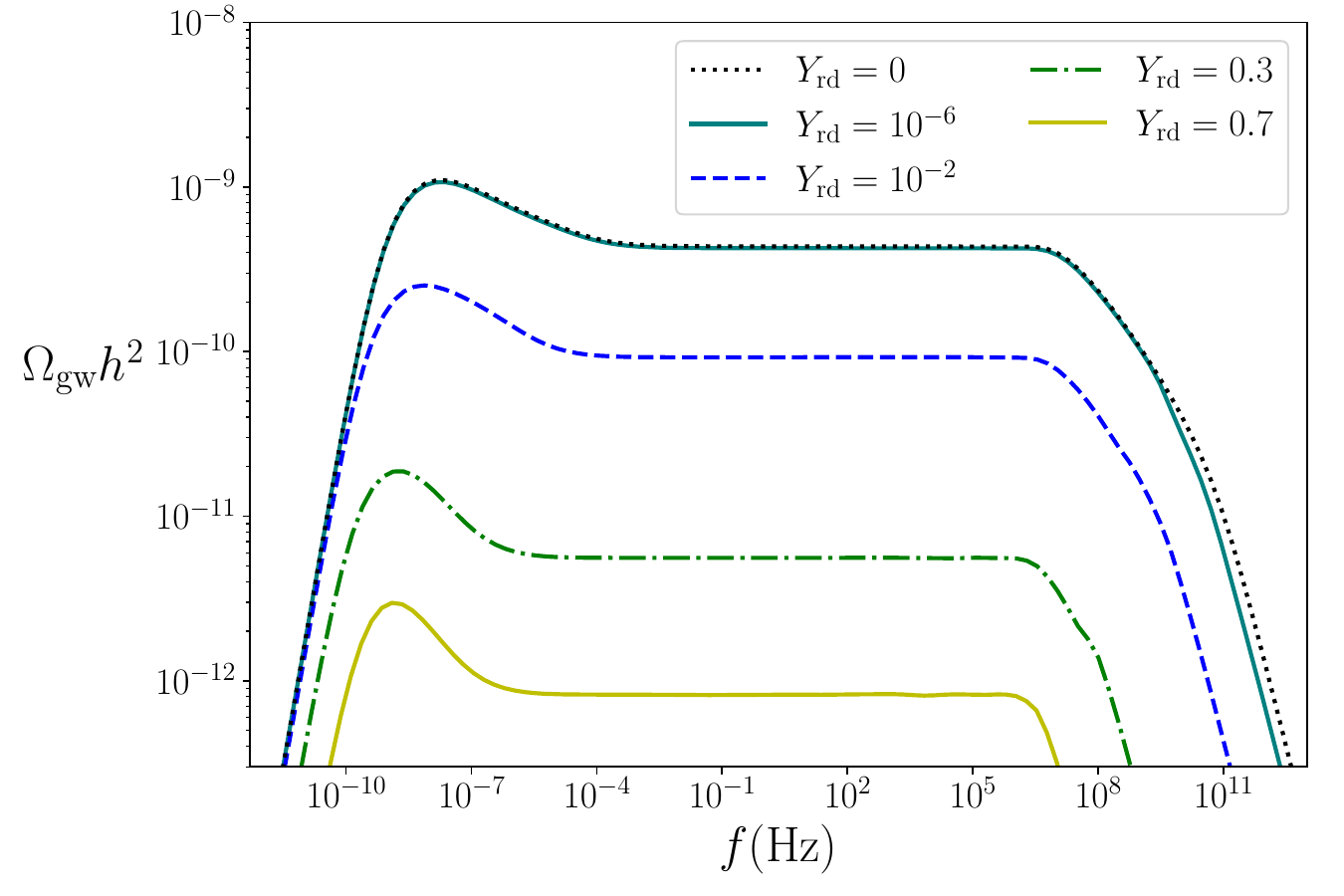}
\includegraphics[scale=0.38]{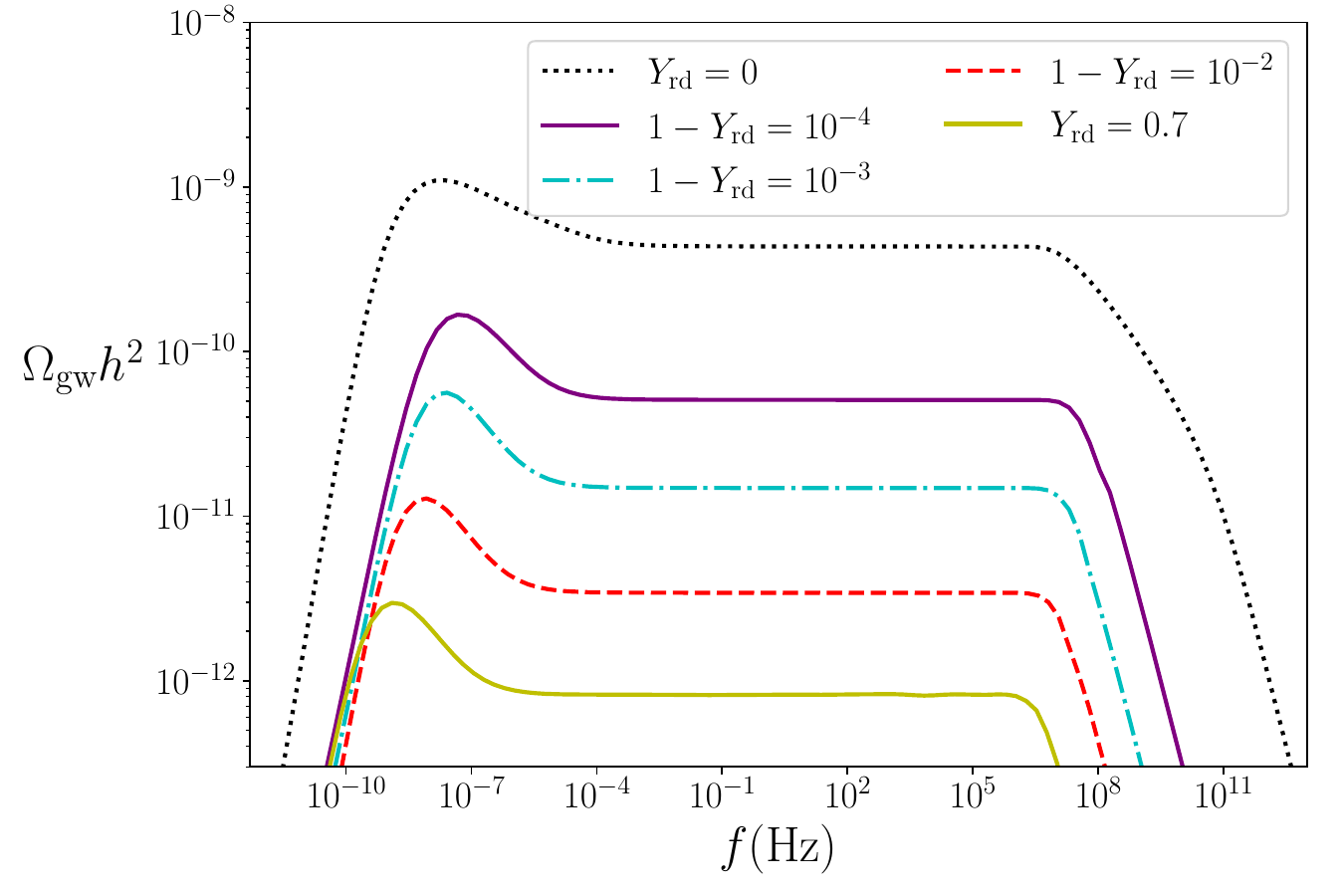}
\caption{\label{Fig:SGWB_cusps} 
The SGWB generated by current-carrying cosmic string networks for different values of $Y_\rd$, for the case in which quasi-cusps dominate the GW emission of loops. The left panel spans the range $0 < Y_{\rm rd} \leq 0.7$, while the right panel represents $0.7 \leq Y_{\rm rd} < 1$. The dotted black line corresponds to the SGWB generated by standard Nambu-Goto string network. All plots are derived for fixed parameters: $G \mu_0 = 10^{-10}$, $\alpha=0.34$, $\mathcal{F}_{\rm fuzz} = 0.1$, $\tilde{e} = 10^{-4}$.}
\end{center}
\end{figure*}

Let us now consider the overall impact of current on the SGWB generated by superconducting strings. Throughout this section, and for the remainder of this paper, for simplicity we will resort to the approximation in Sec.~\ref{subsec:cvosapp} to compute the SGWB spectrum. In Fig.~\ref{fig:YapprComparison}, we display examples of the SGWB computed using the full CVOS system and this approximation. Therein, one may see that there is an excellent agreement between these two approaches.

As we have seen, current should not only affect the total gravitational wave emission efficiency, but it is also expected to affect the spectrum of emission of gravitational radiation for loops with quasi-cusps. If the GW emission of loops is dominated by the emission of kinks, the spectrum of emission reduces to a power law (cf. Eqs.~\eqref{PowerRadGr} and~\eqref{gCK}). In this case, the emission spectrum is similar to that of bare cosmic string loops, but with a different gravitational radiation emission efficiency $\ggr=\ggr_0(1-\yl^{1/2})^B$, and we can factor $j^{-q}$ out from the integration. Thus, one obtains
\begin{equation}
\label{FullOmegaK}
    \Omega_{\rm gw}^{\rm kink}(f)=\frac{16\pi}{3}\left(\frac{G\mu_0}{H_0}\right)^2 \ggr_0 \sum^{n^*}_{j=1}  \frac{j^{-q}}{\mathcal{E}}  \tilde{\Omega}_{\text{gw} \; j }^{\rm{kink} }(f)\,,
\end{equation}
where 
\begin{equation}
\begin{gathered}
\label{OmegaJK}
\tilde{\Omega}_{\rm gw \; \textit{j}}^{\rm kink }(f)= \\
= \frac{j}{f} \int_{t_i}^{t_0}  \left(\frac{a(t)}{a_0}\right)^5 (1-\yl^{1/2})^B n(\ell_j(t'),t')  d t'\,.
\end{gathered}
\end{equation}
In this case, we have that~\cite{SousaAvelino2}
\begin{equation}
    {\tilde{\Omega}_{\text{gw} \; j}^{\rm {kink} }}(f) = {\tilde{\Omega}_{\text{gw} \; j/\beta }^{\rm kink }}(f)( f / \beta ) 
\end{equation}
(as may be seen from Eq.~\eqref{LoopLength2}), which helps us reduce computational time when calculating the sum over harmonic modes of emission as in this case one only needs to compute the contribution of the fundamental mode of emission.

In Fig.~\ref{Fig:SGWB_kinks}, we display the SGWB generated by cosmic string loops with kinks for different values of $Y_{\rm rd}$ for $\te=10^{-4}$ (a regime in which both vector and gravitational radiation have a relevant impact on the evolution of cosmic string loops). In the left panel, we plot the results obtained for string networks with $Y_{\rm rd} \in [0,0.7]$. For this range of values, it may be observed that an increase in the current amplitude leads to a reduction in the amplitude of the SGWB. This reduction occurs due to the additional channel of radiation for loops associated with the vector field. As $Y_{\rm rd}$ increases, more of the loops' energy is lost in the form of vector radiation rather than gravitational radiation. The change in $Y_{\rm rd}$ also affects the evolution of the string network: an increase of the current amplitude in the range $0<Y_{\rm rd}<0.7$ makes the network less dense and slower, as  shown in the right panel of Fig.~\ref{fig:CVOS2}, which translates to a reduction of the number of loops produced. These factors collectively contribute to a general reduction in the SGWB magnitude.

When the current amplitude exceeds $Y_{\rm rd} \sim 0.7$, the magnitude of the SGWB exhibits a notable escalation as $Y_{\rm rd}$ approaches unity, as illustrated in the right panel of Fig.~\ref{Fig:SGWB_kinks}. This phenomenon arises due to the pronounced decrease in gravitational radiation emission efficiency $\ggr$ for high current amplitudes, coupled to a significant enhancement in the production of loops as the network becomes significantly denser (cf. right panel of Fig.~\ref{fig:CVOS2}). As a result, numerous string loops persist to the present epoch (since they decay more slowly) and the GWs they emit are less diluted by expansion. This, coupled to the enhancement caused by the decrease in the frequency of emission, contributes to the amplification of the magnitude of the SGWB. Although the shape of the spectrum remains similar to that of currentless strings, the displacement of the low-frequency peak toward higher frequencies --- which is more evident for large currents --- serves as a signature to distinguish between these two scenarios.

The picture is similar when quasi-cusps dominate the GW emission of current-carrying loops. Current, however, in this case also causes an exponential suppression of the radiation for higher harmonic modes, as it is stated in Eq.~\eqref{PowerRadGr}. We then cannot use relations \eqref{FullOmegaK} and \eqref{OmegaJK} and need to perform an explicit computation for the contribution of all harmonics due to the non-trivial behavior of the function $g(\mathcal{Y})$ in Eq.~\eqref{gCK}\footnote{Note, however, that since $\ell_j = j\ell_1$, $n(\ell_j(t), t)$ does not need to be computed separately for each harmonic mode; only the multiplicative factors in the integral differ.}. We display the SGWB generated by loops with quasi-cusps for different values of the current amplitude $Y_{\rm rd}$ in Fig.~\ref{Fig:SGWB_cusps}. In the left panel, one may see that the amplitude of the SGWB decreases when the current amplitude lies within the interval $0 < Y_{\rm rd} < 0.7$, as was the case for loops with kinks. Conversely, for larger current amplitudes, the amplitude increases steeply and the shift toward higher frequencies of the spectrum becomes more evident as well.

\begin{figure}[t]
\begin{center}
\includegraphics[width=1\linewidth]{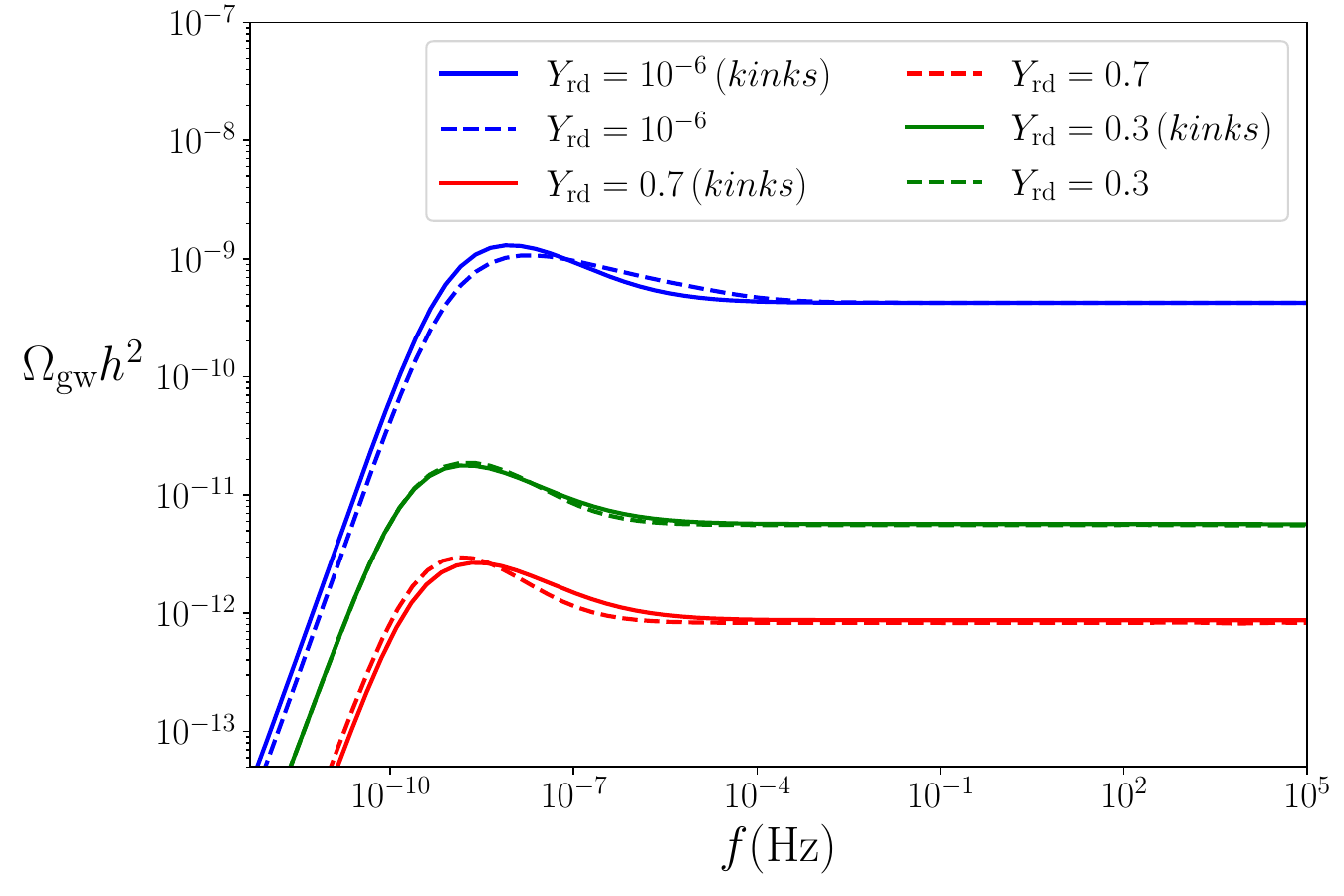}
\caption{\label{Fig:Kinks-Cusps}SGWB generated by a network with only kinks (solid lines) or by only quasi-cusps (dashed lines for different values of $Y$. All plots are derived for fixed parameters: $G \mu_0 = 10^{-10}$, $\alpha=0.34$, $\mathcal{F}_{\rm fuzz} = 0.1$, $\tilde{e} = 10^{-4}$.}
\end{center}
\end{figure}

We also display the SGWB generated by loops with kinks alongside that generated by loops with quasi-cusps in Fig.~\ref{Fig:Kinks-Cusps}. For a small charge amplitude of $Y_{\rm rd} \sim 10^{-6}$, the low-frequency peak for quasi-cusps is shifted to higher frequencies compared to the peak generated by loops with kinks; a similar trend is observed for currentless strings. However, as the current increases, higher harmonic modes are increasingly suppressed due to the exponent in Eq.~\eqref{PowerRadGr}. In fact, for $Y_{\rm rd}=0.3$, we see that both spectra roughly coincide, and for $Y_{\rm rd}=0.7$, the peak of the spectrum generated by loops with kinks actually appears at higher frequencies. The shape, location, amplitude, and broadness of the peak generated by loops with quasi-cusps are then highly dependent on the amplitude of the current.

\section{Cosmological implications and Conclusions}

We have developed methods to compute the SGWB generated by current-carrying cosmic string loops, including the effects of a potential additional decay channel: the emission of vector field radiation. We studied the vector radiation generated by Burden loops and by loops with kinks and demonstrated that the spectrum of emission of kinky loops follows a power law, while that of Burden loops is exponentially suppressed with increasing harmonic mode. We have established a phenomenological relation (Eq. \eqref{Gfun}) that provides a good description of the efficiency of emission of vector field radiation for both these types of loops. The results obtained for currents of the chiral and transonic type are similar, which seems to suggest that the relation between the vector radiation emission efficiency and current obtained may be valid for any other type of current.

\begin{figure*}[t]
\begin{center}
\includegraphics[scale=0.42]{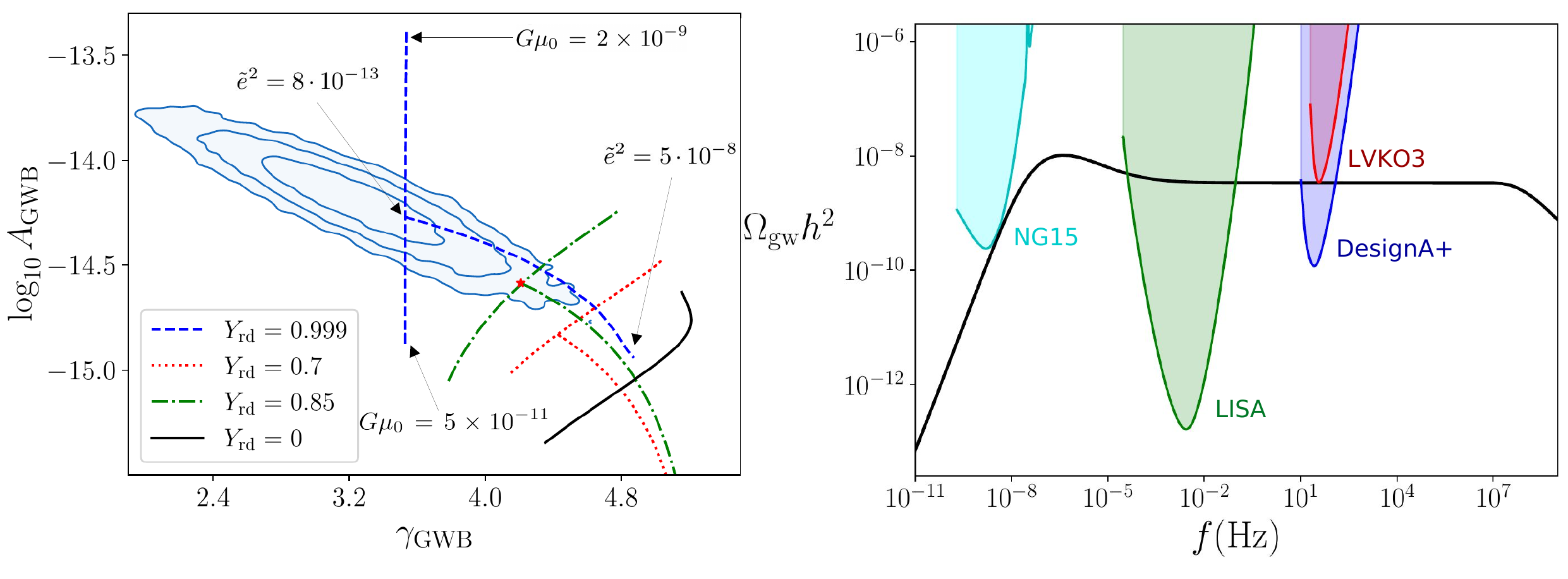}
\caption{\label{Fig:NANOgrav} The left panel depicts the probability distribution of the spectral exponent $\gamma_{\rm GWB}$ and SGWB amplitude $A_{\rm GWB}$ obtained by NANOGrav \cite{NANOGrav:2023gor} for $f_{\rm ref} = 32\text{nHz}$ and $f_{\rm yr} = 1 \text{yr}^{-1}$. Dashed lines represent values of $\gamma_{\rm GWB}$ and $A_{\rm GWB}$ for string tension $G \mu_0$ ranging from $ 2 \times 10^{-9}$ to $ 5 \times 10^{-11}$ and coupling $\tilde{e}^2$ ranging from $8 \times 10^{-13}$ to $5 \times 10^{-8}$, with $Y_{\rm rd} = 0.999$. A dash-dotted line denotes $Y_{\rm rd} = 0.85$, dotted line indicates $Y_{\rm rd} = 0.7$ and solid line corresponds to Nambu-Goto case. The red star represents a specific realization of a current-carrying string network with $G \mu_0 = 2 \times 10^{-10}$, $\tilde{e}=0$, and $Y_{\rm rd} = 0.85$.
The right panel illustrates the spectrum of this specific realization of a string network alongside the LIGO-Virgo-KAGRA O3 constraints and the designed $A+$ sensitivity curve \cite{KAGRA:2021kbb}, as well as the sensitivity curve of a future mission LISA and NANOGrav 15 yr \cite{NANOGrav:2023ctt}. }
\end{center}
\end{figure*}

 To characterize the evolution of current-carrying cosmic string networks and the number density of loops produced, we have resorted to the CVOS model. Using our results for vector radiation, we developed a formalism to compute the spectral density of the SGWB generated by current-carrying cosmic strings. In addition to the bare string tension, $\mu_0$ and the loop size and fuzziness parameters, describing the impact of current and vector radiation requires the inclusion of at least two additional parameters: the current amplitude during the radiation-dominated era $Y_{\rm rd}$ and the charge of current carriers $\tilde{e}$. We have studied the impact of vector radiation on the shape and the amplitude of the SGWB in detail. We demonstrated that when cosmic strings have a strong coupling with the vector fields --- i.e., when $\te^2 \gg G \mu_0$ --- the dominant decay channel of loops is the emission of vector radiation. In this scenario, the amplitude of the SGWB is suppressed due to this additional energy loss channel, but the spectrum may have distinct signatures that allow us to discriminate between this spectrum and that generated by bare cosmic strings. Superconducting cosmic strings may then evade the stringent constraints that result from current GW data, in this limit. In particular, if the vector field to which strings are coupled is standard electromagnetism, we found that this suppression of the amplitude is, in fact, very significant and this type of string should be out of reach of GW detectors, but these scenarios may be constrained also by probing their electromagnetic counterparts \cite{Brandenberger:2019lfm, Imtiaz:2020igv, Cyr:2023yvj}. We also found that the emission of vector radiation may have significant impact on the detectability of the SGWB generated by superconducting strings even for small values of $\te$. As a matter of fact, since the relative importance of vector radiation as a decay channel grows as tension decreases, it will eventually become the dominant decay mechanism for cosmic string loops as one lowers cosmic string tension. Once this regime is reached, the amplitude of the spectrum decreases rapidly as tension is lowered, which necessarily implies that future GW detectors should not be expected to probe current-carrying-string-forming scenarios down to tensions as low as for currentless strings even for small $\te$ values. This is further exacerbated by the fact that, in this regime, the peak of the spectra stops shifting toward higher frequencies as tension decreases (as is the case when GW emission is the dominant decay mechanism of loops). This shift is determinant in LISA's ability to probe Nambu-Goto strings up to tensions as low as $G\mu_0 \sim 10^{-16}-10^{-17}$~\cite{LISA,Blanco-Pillado:2024aca}, as the peak of the spectrum enters the LISA window as tension is lowered. Since the amplitude of the radiation era plateau of the spectrum is significantly lower than the amplitude of the peak, LISA should have less sensitivity to current-carrying string scenarios even for small $\te$ (although this may mean that the peak of the spectrum may be accessible to future pulsar-timing-arrays such as the Square Kilometer Array up to lower tensions in the presence of current).

We also studied how the variation of the charge amplitude $Y_{\rm rd}$ affects the SGWB prediction from superconducting cosmic string networks. We found two regions of $Y_{\rm rd}$ values in which the amplitude of the SGWB has a different behaviour, determined by the distinct impact current has on the number of loops produced in these distinct ranges. The first region, $0 < Y_{\rm rd} < 0.7$, is characterized by a general decrease of the amplitude of the SGWB. In the second region, $0.7 < Y_{\rm rd} < 1$, there is an enhancement of its amplitude and a shift of the spectrum toward higher frequencies (which is always present, but is more evident when current is large). As a matter of fact, as was also found in~\cite{Paper1}, as current approaches its maximum value, this amplitude grows very steeply, especially in the regime in which the emission of GWs is the dominant decay channel for loops.

We conclude our work with some cosmological implications of superconducting cosmic string networks. Although when vector radiation is an important decay channel current-carrying strings may evade current constraints that result from GW data, it is interesting to note that for small enough $\te$, current may actually make their SGWB more compatible with the signal that was recently reported by NANOGrav~\cite{NANOGrav:2023gor}, especially if current is relatively large.  By calculating the spectral exponent $$\gamma_{\rm GWB} = 5 - \left. \frac{d \log \Omega_{\rm gw}}{d \log f} \right\rvert_{\rm f=f_{\rm ref}} $$ and the strain amplitude $$A_{\rm GWB} = \sqrt{\frac{3 H_0^2}{2 \pi^2} \frac{\Omega_{\rm gw}}{f^2_{\rm yr}} \left( \frac{f_{\rm yr}}{f_{\rm ref}} \right)^{5-\gamma_{\rm GWB}} }$$ with $f_{\rm ref} = 32 \text{nHz}$ and $f_{\rm yr} = 1 \text{yr}^{-1}$ \cite{NANOGrav:2023gor}, we demonstrate in Fig.~\ref{Fig:NANOgrav} that the SGWB generated by superconducting cosmic string networks can, in principle, be brought into agreement with NANOGrav pulsar timing data \cite{NANOGrav:2023gor}. As a matter of fact, in the right panel of Fig.~\ref{Fig:NANOgrav}, we present an example of a superconducting string configuration that aligns more closely than Nambu-Goto strings with the NANOGrav data while also avoiding current LIGO-Virgo-KAGRA (LVK) constraints. A similar result was previously explored in Ref.\cite{Afzal:2023kqs} for metastable superconducting strings using a model without vector radiation. Note, however, that a more detailed analysis is needed to determine whether a superconducting cosmic string network can serve as a viable explanation for the timing pulsar data. In particular, the SGWB from the superconducting cosmic string models that remain within $1 \sigma$ of the NANOGrav data seems to be inconsistent with the LVK O3 constraints \cite{KAGRA:2021kbb} and may also violate CMB constraints \cite{Rybak:2024djq}.

We need to emphasize that further study of superconducting string dynamics is required for more precise predictions of their SGWB. In particular, it is important to have a better understanding of vector radiation back-reaction \cite{Amsterdamski:1989xd, Gangui:1997bi, Wachter:2014mja} and the evolution of current-carrying loops via simulations of infinitely thin \cite{Martin:2000ca, Cordero-Cid:2002hmv} and field-theory cosmic string loops \cite{Battye:2021sji}. We also did not study the possible effect on SGWB caused by the formation of vortons \cite{Auclair:2020wse}. We leave these problems for our future investigations.

\acknowledgments
I.R. would like to express gratitude to the organizers of the Workshop on Topological Defects in Daejeon at the Institute for Basic Science CTPU-CGA for their helpful and enlightening discussions (Christophe Ringeval, Mairi Sakellariadou, Tanmay Vachaspati, and Masahide Yamaguchi). Additionally, I.R. is thankful for the valuable conversations and important remarks from Bryce Cyr, Pierre Auclair, and Tanmay Vachaspati.

This work was financed by Portuguese funds through FCT -
Funda\c{}c\~ao para a Ci\^encia e a Tecnologia in the framework of the R\&D project 2022.03495.PTDC - \textit{Uncovering the nature of
cosmic strings}. I.R. also acknowledges support from the Grant PGC2022-126078NB-C21 funded by MCIN/AEI/ 10.13039/501100011033 and ``ERDF A way of making Europe'', as well as Grant DGA-FSE grant 2020-E21-17R from the Aragon Government and the European Union - NextGenerationEU Recovery and Resilience Program on `Astrof\'{\i}sica y F\'{\i}sica de Altas Energ\'{\i}as' CEFCA-CAPA-ITAINNOVA. L. S. was also supported by FCT through contract No. DL 57/2016/CP1364/CT0001 and through the research grants UIDB/04434/2020 and UIDP/04434/2020.

\appendix
\begin{widetext}
\section{Emission of vector radiation by Burden loop}
\label{Burden's loop}

To study the emission of vector radiation by loops with quasi-cusp points, we will resort to a particular solution proposed by Burden in~\cite{BURDEN1985} (illustrated in the left panel of Fig.~\ref{fig:TwoLoopsExample})
\begin{equation}
\begin{gathered}
\label{ABmodesLoops1}
\textbf{X}_- = \ell G_- \left[ \frac{\textbf{e}_3}{N_-} \cos \frac{\sigma_- N_-}{\ell} + \frac{ \textbf{e}_1}{ N_-} \sin \frac{ \sigma_- N_-}{\ell} \right] , \\
\textbf{X}_+ = \ell G_+ \Bigg[ \frac{\textbf{e}_3}{N_+} \cos \frac{ N_+ \sigma_+ }{\ell}  + \frac{\sin \frac{N_+ \sigma_+}{\ell} }{ N_+} \left( \textbf{e}_1 \cos \beta + \textbf{e}_2 \sin \beta  \right)   \Bigg],
\end{gathered}
\end{equation}
where $\sigma_{\pm} \in [0, \, 2 \pi ]$, $G_{\pm} = \sqrt{ 1 - F^{\prime \, 2}_{\pm}(\sigma_{\pm})}$ are treated as constants,  $\textbf{e}_i$ are orthonormal vectors and $N_\pm$ are integers that are relatively prime. This solution describes loops with trajectories that do not self-intersect and that form quasi-cusps at discrete instants of time.

For this loop solution, the integrals in Eq.~(\ref{IJpm}) may be expressed in terms of Bessel functions of the first kind $J_n (\dots)$ and their derivative $J^{\prime}_n (\dots)$ as

\begin{equation}
\begin{gathered}
    \label{PnFinalBurd}
   \frac{ d \gem_j }{d \Omega} = 2 \pi j^2 J_{p +}^2 J_{p-}^2 \times \\
    \Bigg[ \mathcal{J}_{+} + \mathcal{J}_{-} 
    + 2 F_+^{\prime} F_-^{\prime} \left( 1 - \frac{G_+}{d_+} \frac{G_-}{d_-} \left( \frac{ b_+ b_- \cos \beta + a_+ a_- }{d_- d_+} + \frac{ J_{+}^{\prime} J_{-}^{\prime} \left( b_+ b_- + a_- a_+ \cos \beta \right) }{J_{+} J_{-}}  \right) \right) \Bigg]\,,
\end{gathered}
\end{equation}
where we have defined
\begin{equation}
       \mathcal{J}_{\pm} = F_{\mp}^2  G_{\pm}^2 \left( \frac{1}{d_{\pm}^2} + \frac{J_{p_{\pm}}^{\prime\, 2 }}{J_{\pm}^2}   \right) - F_{\mp}^2\,,\quad
       d_{\pm} = b_{\pm} \sqrt{ 1 + \frac{a_{\pm}^2}{b_{\pm}^2} }\,,
\end{equation}
$J_{\pm} = J_{p_{\pm}}(p_{\pm} d_{\pm})$, $p_{\pm} = {j}/{N_{\pm}}$, $\delta_{\pm} = \arctan \left( {a_{\pm}}/{b_{\pm}} \right)$, $a_{\pm} = F^{\prime}_\pm n_z$, $b_{+} = F^{\prime}_+ n_x $, and $b_- = F^{\prime}_- \left( \cos \beta + \sin \beta \right) $, and we have used the following relations:
\begin{equation}
\begin{gathered}
\label{InegrLoop1}
\int_0^{2 \pi} \text{e}^{i j \left( x - \frac{a}{N} \cos N x - \frac{b}{N} \sin N x \right)} \begin{cases} \sin N x \\ \cos N x \\ 1 \end{cases} d x = 2 \pi \text{e}^{-i n \delta} \begin{cases}
  i J^{\prime}_{n} \cos \delta  -  \frac{ \sin \delta }{d} J_n , \quad \text{if: } \; j \mid N \\
    i J^{\prime}_{n} \sin \delta  +  \frac{ \cos \delta }{d} J_n  , \quad \text{if: } \; j \mid N, \\
   J_n, \quad \text{if: } \; j \mid N.
\end{cases}
\end{gathered}
\end{equation}
where $J_n = J_n (n d)$, $J_n^{\prime} = J^{\prime}_n (n d)$, $n= {j}/{N}$, $\delta = \arctan ({a}/{b})$ and 
$d = b \sqrt{1 + \left( \frac{a}{b} \right)^2}$.  The integral in \eqref{InegrLoop1} vanishes for all other choices of $j$ and $N$ (see Appendix in Ref.~\cite{BabichevDokuchaev2002} for more detail).

By integrating Eq.~\eqref{PnFinalBurd} over the solid angle $\Omega$, we may obtain the vector radiation emission efficiency, in each harmonic mode $j$, $\gem_j$ and, by summing over all harmonic modes, we may obtain the total vector radiation emission efficiency $\gem$. The results we obtain are displayed in Fig.~\ref{fig:BurdenGe}. Therein, on the left panel, one may see that $\gem_j$ decays exponentially with increasing harmonic mode and, as a result, higher harmonics provide a negligible contribution to $\gem$ and therefore the summation in Eq.~\eqref{PowerRad} may safely be truncated. We found that it is sufficient to include the first 150 harmonic modes in the computation of $\gem$ (right panel). In fact, as Fig.~\ref{fig:BurdenExp} shows, the spectrum of emission is very well described by a function of the form $\gem_j \propto \text{e}^{- j f_m(G_{\pm})}$ for both chiral and symmetrical currents.

\begin{figure}[t]
\centering	\includegraphics[scale=0.55]{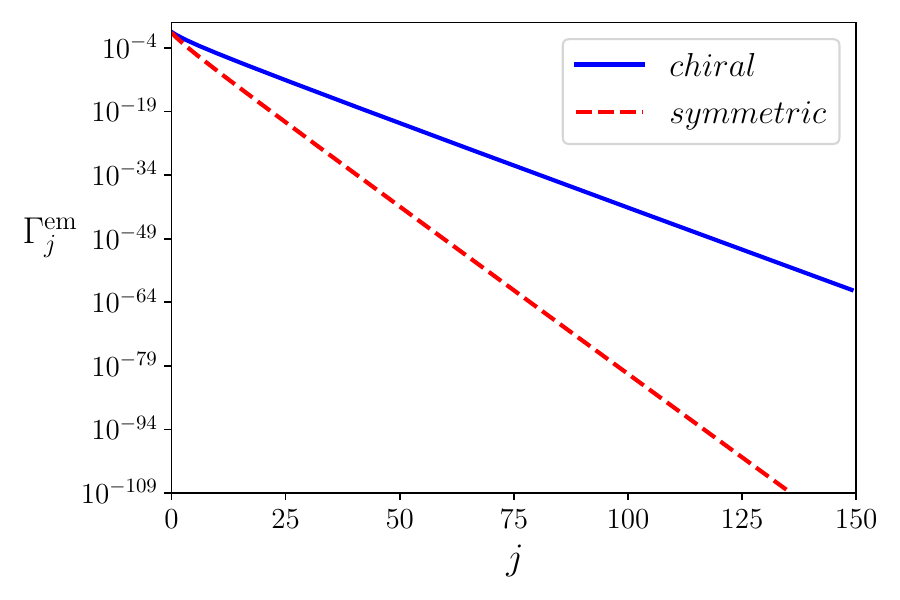}
  \includegraphics[scale=0.55]{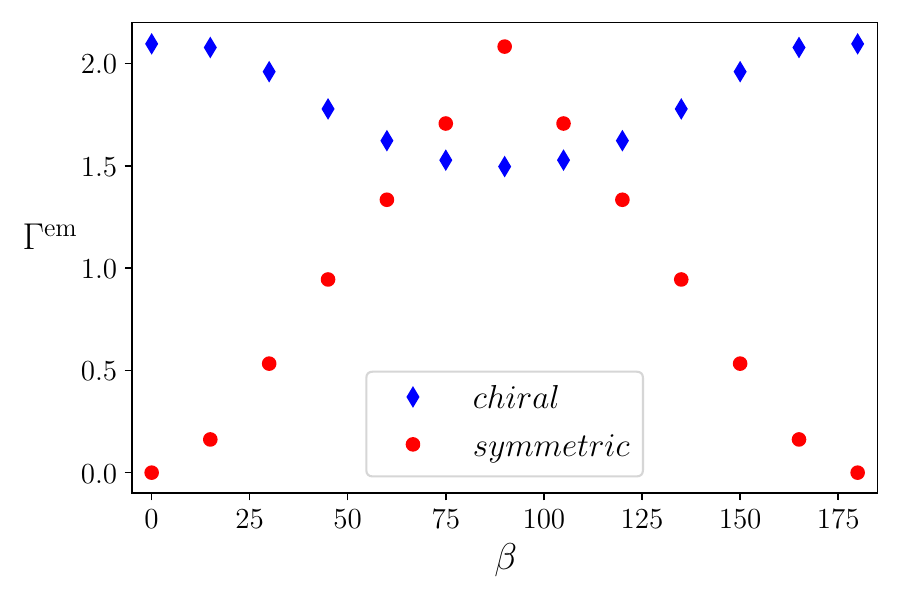}
\caption{ \label{fig:BurdenGe}  Vector radiation emission efficiency of loops with quasi-cusps. The left panel displays the emission spectrum of vector radiation $\gem$ for Burden loops with $\beta={\pi}/{2}$. The solid line corresponds to chiral loops with $G_{+}=1$ and $G_{-}=0.5$, while the dashed line representsloops with a symmetrical current given by $G_{\pm}=0.5$. 
\newline 
The right panel displays the vector radiation emission efficiency as a function of $\beta$ and $\gem$,  for Burden loops, constructed from the first $150$ harmonic modes. Diamonds (blue) represent the values of $\gem$ for chiral loops with $G_{+}=1$ and $G_-=0.9$, while the circles (red) correspond to loops with symmetrical currents ($G_{\pm}=0.9$). All results were obtained for $N_{\pm}=1$. }
\end{figure}

\begin{figure}[t]
  \includegraphics[scale=0.6]{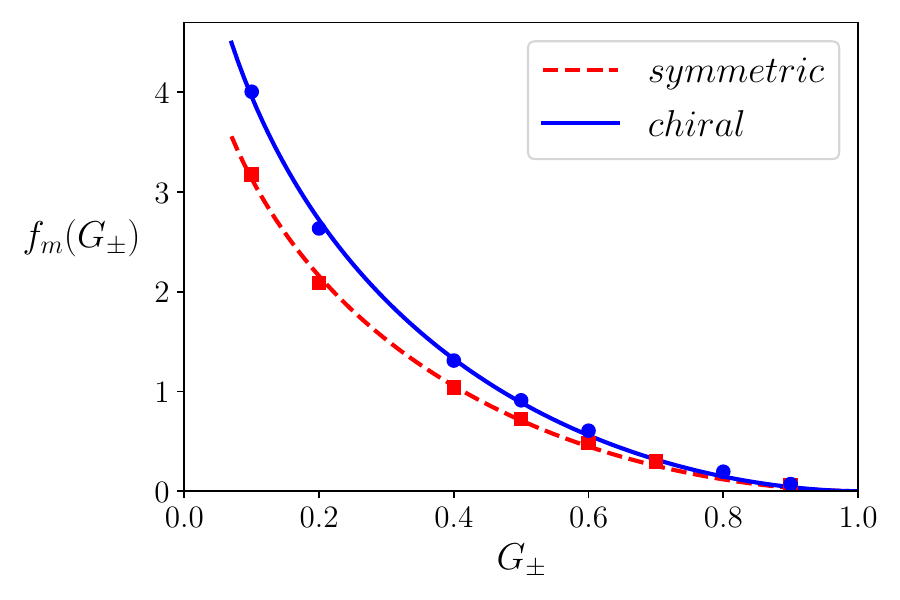}
  \includegraphics[scale=0.28]{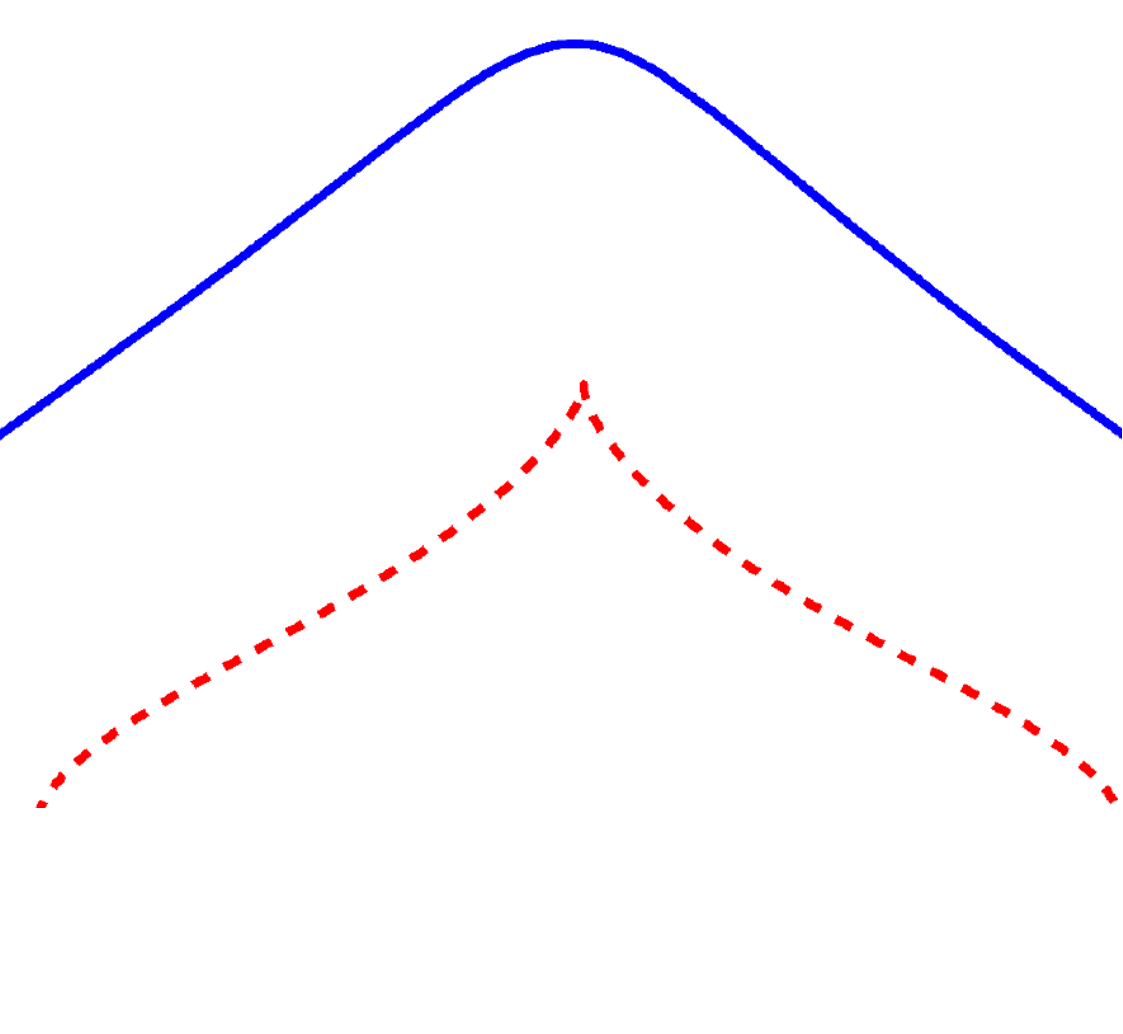}
\caption{ \label{fig:BurdenExp} The left panel displays the fitting of the spectrum of emission of vector radiation by loops with quasi-cusps by an exponentially decaying function of the form $\gem_j \propto \text{e}^{- j f_m(G_{\pm})}$, with $f_{m}(G_{\pm}) = a_m \left( 1 - \sqrt{G_{\pm}}  \right)^{b_{m}} $. Red dots (or blue diamonds) represent the values computed for $\gem_j$ for various values of $G_\pm$ for loops with chiral (or symmetrical) currents, while the solid (dashed) line represents the best fit values for $f_m(G_{\pm})$ in each case respectively. The fitting constants are, for symmetrical current, given by $a_s = 6.01$, $b_{s}=1.75$, while for chiral loops we found $a_{c} = 7.73$, $b_{c}=1.76$ cases. \newline
The right panel illustrates an example of a quasi-cusp shape on Burden loops ($N_+ = 3$, $N_- = 1$) for both symmetric ($G_{\pm} = 0.5$, shown as a dashed line) and chiral ($G_+ = 1$, $G_- = 0.5$, shown as a solid line) currents. The shape in case of a symmetric current matches the Nambu-Goto shape but moves at a subluminal velocity, whereas, for the chiral current, the quasi-cusp shape appears smoothed.}
\end{figure}

\section{Emission of vector radiation by loops with kinks}
\label{Cuspless loop}

To study the emission of vector radiation by kinks, one may consider a class of loops consisting of 4 straight segments that was introduced originally in Refs.~\cite{GarfinkleVachaspati2, GarfinkleVachaspati, CopelandHindmarshTurok} and (illustrated in the right panel of Fig.~\ref{fig:TwoLoopsExample}),
\begin{equation}
\label{CuspLessLoop}
\textbf{X}_{\pm} = G_{\pm} \textbf{r}_{\pm} \begin{cases} \frac{\ell}{2 \pi} \sigma_{\pm} - \frac{\ell}{4}, \qquad 0 \leq \sigma_{\pm} \leq \pi, \\ \frac{3 \ell}{4} - \frac{\ell}{2 \pi} \sigma_{\pm}, \qquad \pi \leq \sigma_{\pm} \leq 2 \pi,  \end{cases}
\end{equation}
where we have defined the unit vectors $\textbf{r}_{\pm} = \left( 0, \; \cos \frac{\beta}{2}, \; \mp \sin \frac{\beta}{2} \right)$, with $\beta$ being the angle between these vectors. Note that \eqref{CuspLessLoop} is a solution for the general equation of motion for superconducting strings, characterized by any equation of state, and not just for transonic superconducting strings. This particular solution was, in fact, used to study the vector radiation for loops with chiral currents in Refs.~\cite{BlancoPilladoOlum, BabichevDokuchaev2002}.

By plugging expression (\ref{CuspLessLoop}) into (\ref{IJpm}) and (\ref{PowerRad}), one may carry out the integration to obtain
\begin{figure}[t]
\centering
\includegraphics[scale=0.55]{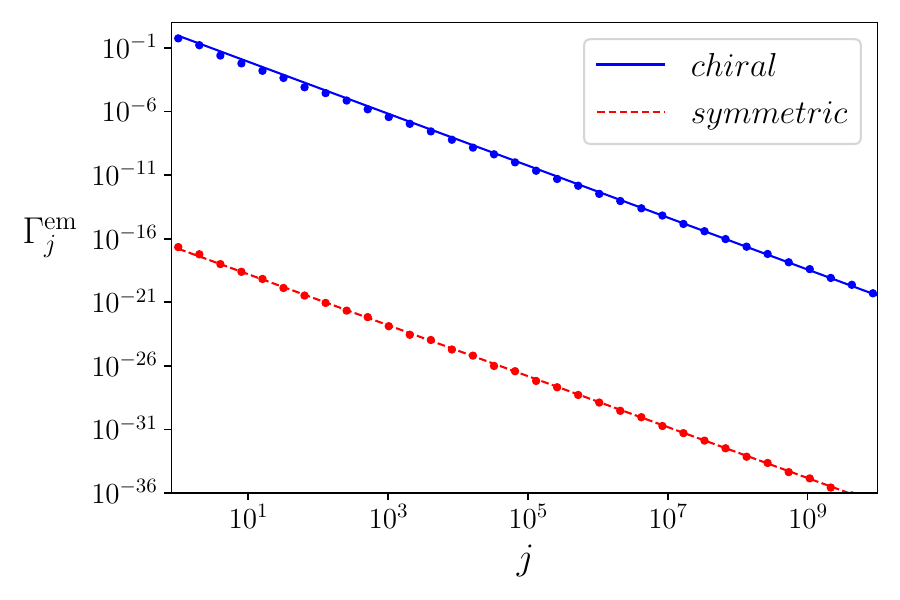}
\includegraphics[scale=0.55]{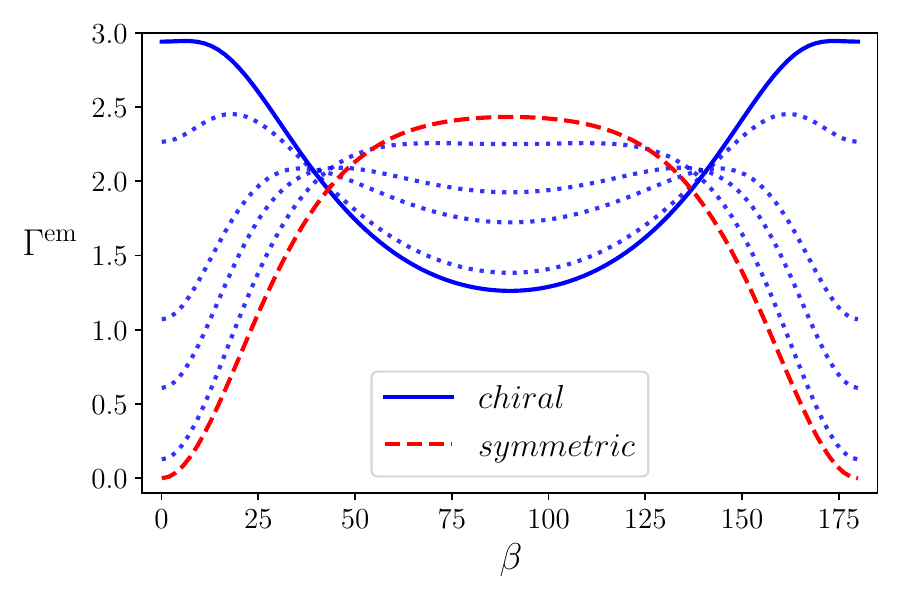}
\caption{\label{fig:CuspLessGe} Vector radiation emission efficiency of Garfinkle-Vachaspati loops. The left panel shows the that the spectrum of emission of vector radiation follows a power law of the form $\gem_j \sim j^{-2}$ for when $\beta={\pi}/{2}$. The solid line corresponds to $G_{+}=1$ and $G_{-}=0.5$, while the dashed line corresponds to $G_{\pm}=0.5$.
 The right panel represents the vector radiation emission efficiency $\gem$ for loops with kinks. The solid line corresponds to a chiral loop with $G_{+}=1$ and $G_-=0.9$, while the dashed line corresponds to a symmetric current loop with $G_{\pm}=0.9$. Dotted lines demonstrate the effect of varying $G_{-} \in (0.9, 1.0)$. }
\end{figure}

\begin{figure}[t]
\centering
\includegraphics[scale=0.6]{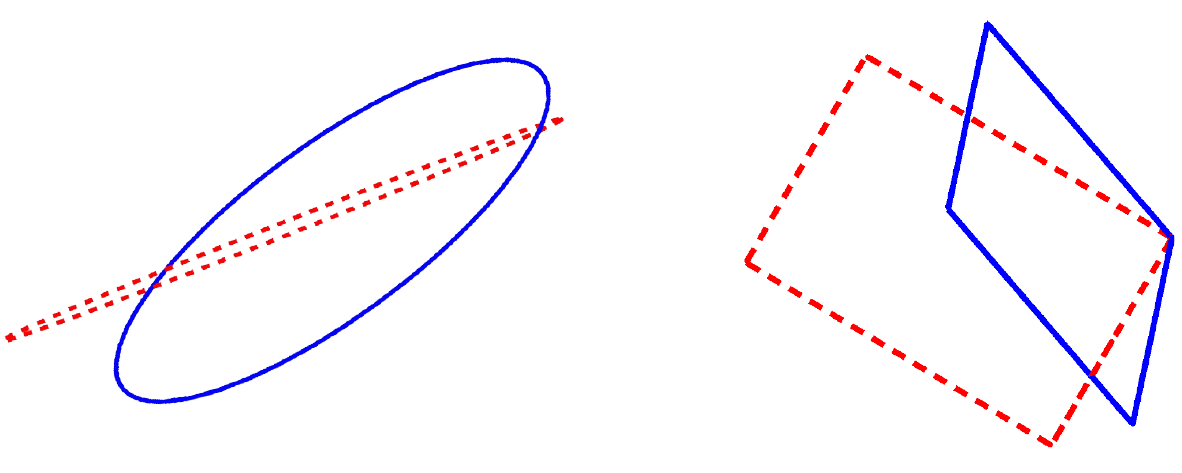}
\caption{\label{fig:TwoLoopsExample} Left panel displays an example of a Burden loop, given by Eq.~\eqref{ABmodesLoops1}, with $N_{\pm}=1$, where the dashed line represents the case in which current is symmetric, defined by $G_{\pm}=1$, and the solid line represents a chiral current with $G_+ = 1$, $G_- = 0.5$. \newline
Right panel shows a moment in the evolution of Garfinkle-Vachaspati loops, given by Eq.~\eqref{CuspLessLoop}, for a symmetric current ($G_{\pm}=1$), depicted by the dashed line, and a chiral current ($G_+ = 1$, $G_- = 0.5$), depicted by the solid line.}
\end{figure}

\begin{equation}
\begin{gathered}
    \label{PnFinalGarf}
    \frac{ d \gem_j }{d \Omega} = \frac{8 G_-^2 G_+^2}{\pi^3}  \frac{\left(1 - (-1)^j \cos(j \pi G_- \textbf{n} \cdot \textbf{r}_-) \right) \left(1 - (-1)^j \cos(j \pi G_+ \textbf{n} \cdot \textbf{r}_+) \right) }{ j^2 \left(1 - G_-^2 (\textbf{n} \cdot \textbf{r}_-)^2 \right)^2 \left(1 - G_+^2 (\textbf{n} \cdot \textbf{r}_+)^2 \right)^2} ( \textbf{n} \cdot \textbf{r}_+ )^2 ( \textbf{n} \cdot \textbf{r}_- )^2 \times \\
    \left[ F^{\prime \, 2}_+ \frac{ 1 - ( \textbf{n} \cdot \textbf{r}_- )^2 }{ ( \textbf{n} \cdot \textbf{r}_- )^2 } + F^{\prime \, 2}_- \frac{ 1 - ( \textbf{n} \cdot \textbf{r}_+ )^2 }{ ( \textbf{n} \cdot \textbf{r}_+ )^2 } + 2 F^{\prime}_+ F^{\prime}_- \frac{ ( \textbf{n} \cdot \textbf{r}_- ) ( \textbf{n} \cdot \textbf{r}_+ ) - \textbf{r}_- \cdot \textbf{r}_+ }{ ( \textbf{n} \cdot \textbf{r}_+ ) ( \textbf{n} \cdot \textbf{r}_- ) } \right],
\end{gathered}
\end{equation}
\begin{equation}
\begin{gathered}
    \label{PnFinalGarf2}
  \frac{ d \gem }{d \Omega} = \frac{4 G_-^2 G_+^2}{\pi}  \frac{\left( 1 - \frac{1}{2} \left( | G_+ \textbf{n} \cdot \textbf{r}_+ + G_- \textbf{n} \cdot \textbf{r}_- | +  | G_+ \textbf{n} \cdot \textbf{r}_+ - G_- \textbf{n} \cdot \textbf{r}_- | \right) \right) }{ \left(1 - G_-^2 (\textbf{n} \cdot \textbf{r}_-)^2 \right)^2 \left(1 - G_+^2 (\textbf{n} \cdot \textbf{r}_+)^2 \right)^2} ( \textbf{n} \cdot \textbf{r}_+ )^2 ( \textbf{n} \cdot \textbf{r}_- )^2 \times \\
    \left[ F^{\prime \, 2}_+ \frac{ 1 - ( \textbf{n} \cdot \textbf{r}_- )^2 }{ ( \textbf{n} \cdot \textbf{r}_- )^2 } + F^{\prime \, 2}_- \frac{ 1 - ( \textbf{n} \cdot \textbf{r}_+ )^2 }{ ( \textbf{n} \cdot \textbf{r}_+ )^2 } + 2 F^{\prime}_+ F^{\prime}_- \frac{ ( \textbf{n} \cdot \textbf{r}_- ) ( \textbf{n} \cdot \textbf{r}_+ ) - \textbf{r}_- \cdot \textbf{r}_+ }{ ( \textbf{n} \cdot \textbf{r}_+ ) ( \textbf{n} \cdot \textbf{r}_- ) } \right],
\end{gathered}
\end{equation}
which reduce to the results obtained in Ref.~\cite{BabichevDokuchaev2002} for chiral loops if one sets $F^{\prime}_- = 0$ (or $F_+^{\prime}=0$).

In Fig.~\ref{fig:CuspLessGe}, we display the resulting spectrum of emission of vector radiation for Garfinkle-Vachaspati loops $\gem_j$ and their vector radiation emission efficiency $\gem$ (obtained, respectively after carrying out an integration over solid angle and summing over harmonic modes). Therein, one may see that, in this case, the spectrum behaves as a power law of the form $\gem_j\propto j^{-2}$ and there is no additional exponential suppression with increasing $j$ as we have seen in the case of loops with cusps. A similar picture was found for the GW emission power spectrum in~\cite{Paper1}.

\end{widetext}

\section{No cusps for current-carrying strings}
\label{NoCusps}

A variation of the action \eqref{EffectiveAction} with respect to $X^{\mu}$ and $\phi$ yields the following equations of motion:

\begin{equation}
\begin{gathered}
\label{XPhiEq}
\mu_0 \partial_a \left[ \mathcal{T}^{ab} X^{\mu}_{,b} \right] = e F^{\mu}_{\nu} X^{\nu}_{,a} J^a \sqrt{-\gamma}, \\
\mu_0 \partial_a \left[ \sqrt{-\gamma} \gamma^{ab} \phi_{,b} \frac{d\mathcal{L}}{d \kappa} \right] = \frac{e}{4 } \varepsilon^{ac} F_{ac} \sqrt{-\gamma},
\end{gathered}
\end{equation}
where
\begin{equation}
\begin{gathered}
    \mathcal{T}^{ab} = \sqrt{-\gamma} \left( \gamma^{ab} \mathcal{L} - 2 \frac{d\mathcal{L}}{d \kappa} \gamma^{ac} \gamma^{bd} \phi_{,c} \phi_{,d} \right)\,,\\
    J^a= e\frac{\varepsilon^{ab}\phi_{,b}}{\sqrt{-\gamma}}\,,
\end{gathered}
\end{equation}
are, respectively, the stress-energy tensor and the 2-current on the worldsheet and $F_{ac} = X^{\mu}_{,a} X^{\nu}_{,c} F_{\mu \nu}$.

A cusp is defined as a point where induced metric $\gamma_{ab}$ is singular and this string point reaches the speed of light (i.e., $\dot{X}^{ 2} \equiv \left(\frac{d X}{dt} \right)^2 = 0$). Here, we show that having a cusp is not possible when superconductivity is non-vanishing, $\phi_{,a} \neq 0$, at this point. 
If one selects the temporal-transverse gauge for the world-sheet parametrization, as described above Eq.~\eqref{TempGauge}, with $\gamma_{01}=0$ and $\dot{X}^{\mu}$ corresponding the cosmic string's 4-velocity (whose spatial component is perpendicular to the string), the temporal component of the equation of motion \eqref{XPhiEq} in Minkowski space takes the following form:
\begin{equation}
\begin{gathered}
\label{XPhiEq2}
\partial_{t} \left[ \tilde{\varepsilon} \left( \mathcal{L} - 2 \gamma^{00} \dot{\phi}^2 \frac{d\mathcal{L}}{d \kappa}  \right) \right] = - \partial_{s} \left[ \frac{2}{\sqrt{-\gamma}}  \dot{\phi} \phi^{\prime}_s \frac{d\mathcal{L}}{d \kappa} \right] ,
\end{gathered}
\end{equation}
where $\tilde{\varepsilon}^2 =  {-X^{\prime^2}_s}/{\dot{X}^2}$. Once the point on the string reaches the speed of light, it leads to $\gamma_{00} = \dot{X}^2 \rightarrow 0$ and $\gamma^{00} = 1/\dot{X}^2 \rightarrow \infty$; hence, Eq.~(\ref{XPhiEq2}) diverges if $\gamma_{00}\rightarrow 0$ without any other assumptions. To prevent this divergence, it is necessary to demand that $\gamma_{11} \rightarrow 0$ in such a way that $\tilde{\varepsilon} \rightarrow 1$, thereby maintaining the first term free from divergence. However, upon closer examination, it becomes apparent that whenever $\dot{\phi} \neq 0 $ or $\phi^{\prime}_s \neq 0$, the divergence of the second term persists and cannot be compensated by the right-hand side. Consequently, non-trivial superconductivity at a point on the cosmic string worldsheet prevents the formation of cusps regardless of the specific form of the master function $\mathcal{L}(\kappa)$.

\bibliography{Biblio}

\end{document}